\documentclass[aps,superscriptaddress,twocolumn,nofootinbib]{revtex4-2}
\usepackage[utf8]{inputenc} % allow utf-8 input
\usepackage[T1]{fontenc}    % use 8-bit T1 fonts
\usepackage[hidelinks]{hyperref}       % hyperlinks
\usepackage{url}            % simple URL typesetting
\usepackage{booktabs}       % professional-quality tables
\usepackage{nicefrac}       % compact symbols for 1/2, etc.
\usepackage{graphicx}
\usepackage{mathtools}
\usepackage{xcolor}
\usepackage{amsmath}
\usepackage{amssymb}
\usepackage{amsfonts}       % blackboard math symbols
\usepackage{physics}
\usepackage{xspace}
\usepackage{bigints}
\usepackage{appendix}
\usepackage{algorithm}
\usepackage{algpseudocode}
\usepackage[algo2e,ruled,vlined,linesnumbered]{algorithm2e}
\usepackage{mathtools}
\usepackage{paralist}
\usepackage{cleveref}

\newcommand{\Var}{{\rm Var}}

\newcommand{\bigO}{\ensuremath{\mathcal{O}}\xspace}

\newcommand{\vtheta}{\ensuremath{\vec{\theta}}\xspace}

\newcommand{\E}{\ensuremath{\operatorname{\mathbb E}\xspace}}

\newcommand{\epsM}{\ensuremath{\epsilon_{\rm M}}}

\newcommand{\Obs}{\ensuremath{\hat{O}}}

\newcommand{\Normgrad}{\ensuremath{\norm{\nabla C}/C_0}}
\newcommand{\Grad}{\ensuremath{\nabla C}}

\newcommand{\sampcost}{\ensuremath{C}}
\newcommand{\Tone}{\ensuremath{T_{1}}\xspace}
\newcommand{\ToneMean}{\ensuremath{\langle T_{1} \rangle }\xspace}
\newcommand{\effTone}{\ensuremath{T^{{\rm eff}}_{1}}\xspace}
\newcommand{\Ttwo}{\ensuremath{T_{2}}\xspace}

\newcommand{\cirtimes}{t_{cir}\xspace}
\newcommand{\mus}{\ensuremath{\,\mu s}\xspace}
\renewcommand{\trace}{\mathrm{Tr}}

\DeclarePairedDelimiterX{\inp}[2]{\langle}{\rangle}{#1, #2}

\graphicspath{{figures_draft/}} 

\newcommand{\hri}{Honda Research Institute Europe GmbH, Carl-Legien-Str.\ 30, 63073 Offenbach, Germany}
\newcommand{\pitaevskiicenter}{Pitaevskii BEC Center, CNR-INO and Dipartimento di Fisica, Università di Trento, I-38123 Trento, Italy}
\newcommand{\unitrento}{INFN-TIFPA, Trento Institute for Fundamental Physics and Applications, Trento, Italy}

\begin{document}
\title{Experimental demonstration of the absence of noise-induced barren plateaus using information content landscape analysis}

\author{Sebastian Schmitt}
\affiliation{\hri}
\author{Linus Ekstr\o m}
\affiliation{\hri}
\author{Alberto Bottarelli}
\affiliation{\hri}
\affiliation{\pitaevskiicenter}
\affiliation{\unitrento}
\author{Xavier Bonet-Monroig}
\email[]{xavier.bonet@honda-ri.de}
\affiliation{\hri}

\begin{abstract}
Variational quantum algorithms are promising candidates for near-term quantum computing but can be hindered by barren plateaus, where gradients vanish exponentially and optimization becomes intractable.
Noise-Induced Barren Plateaus (NIBP) are particularly concerning because they are predicted to arise generically from noise accumulation, independent of system size, circuit structure, and observable locality.
We experimentally investigate NIBP on IBM quantum hardware. Using Information Content Landscape Analysis (ICLA), we efficiently estimate gradient norms for variational circuits ranging from 8 to 102 qubits, up to hundreds of parameters and circuit runtimes of  hundreds of microseconds.
Contrary to NIBP expectations, we observe that gradient magnitudes saturate beyond a characteristic runtime rather than decaying exponentially.
Classical simulations of the 8-qubit case under noiseless, depolarizing, amplitude-damping, and dephasing noise models support this behavior.
Consistent with recent theory, our results show that $\Tone$-dominated non-unital noise inhibits the emergence of NIBP.
Our analysis suggest that average calibration metrics may be insufficient to predict variational algorithm performance.
\end{abstract}

\maketitle

\section{Introduction}
The significant advances in quantum hardware over the past decade have paved the way toward the pursuit of quantum advantage in the pre-fault-tolerant, or near-term, quantum computing era~\cite{preskill2018quantum}.
The so-called \textit{beyond classical computation} experiments~\cite{arute2019quantum,ZhongGaussianBS2021,WuSupremacy2021} have further fueled the hope that practical quantum advantage could be achieved in the near-term.
However, a decade of research on noisy intermediate scale quantum (NISQ) algorithms~\cite{peruzzo2014variational} has yet to provide a clear path towards such advantage.
In parallel, the progress towards universal Fault-Tolerant Quantum Computers (FTQC) has been immense, reaching important milestones in deploying quantum error correction~\cite{QECSurfaceCodeGoogle2025,innsbruckLogicCodeSwitching2025,queraLogical2023}, lattice surgery for two-qubit logical gates~\cite{LatticeSurgeryETH2026}, and generating magic states~\cite{MagicStateDestiallationQuEra2025,MagicStateCultivationGoogle2025} as resources for non-Clifford gates.
While large-scale deployments of FTQC devices is at a distant future, the size of the Quantum Processing Units (QPU) are large enough to be theoretically hard to simulate, thus bringing a renewed interest in to pre-FTQC algorithms to at least be useful as a scientific discovery tool.

A significant amount of the research in NISQ has been focused on Variational Quantum Algorithms (VQA)~\cite{peruzzo2014variational,cerezo2020variational,bharti2022noisy} due to their flexibility to address a broad class of problems as well as their ability to be tailored to the hardware limitations.
Unfortunately, the requirements for a large-scale deployment of a VQA are considerable e.g.\ the number of quantum state preparations, measurements, and cost function evaluations, or the struggle to handle hardware noise~\cite{bonet-monroig2023performance}.
Barren Plateaus (BP)~\cite{mcclean2018barren,Larocca2025ReviewBP} or vanishing gradients are one of the biggest limitations of VQA as they fundamentally limit the ability  to optimize  a circuit to provide the desired answer.
BP arise for various reasons, such as form of  the cost function~\cite{cerezo2021cost} or number of qubits in the circuit~\cite{Ragone2024LieAlgebraBP}, and emerge with increasing  circuit length.
Of particular interest are the so-called Noise-Induced Barren Plateaus (NIBP)~\cite{WangNIBP2021} which emerge due to  gate errors or depolarizing noise channels, and are of particular importance when transitioning from simulations to hardware experiments 
However, recent theoretical insights~\cite{Pat2025beyondunitalnoise,meleNonunitalNoise2024,feffermanNonUnitalNoisePRXQ2024} have shown that under non-unital noise, the gradient of a VQA does not vanish, and thus there will be no NIBP.
This result also leads to the insight that  NIBP might  never realistically appear because there will always be a dissipative process, typically measured in the form of \(\Tone\) coherence time.
A conclusion that was already stated in ref.~\cite{meleNonunitalNoise2024}.

In this work, we experimentally show the absence of NIBP due to the effect of non-unital noise, characterized by the coherence time for amplitude damping, \(\Tone\).
We run quantum circuits on IBM hardware with system sizes ranging from $N=8$ to $N=102$ qubits and runtimes up to several  hundreds of microseconds.
To efficiently estimate the gradient signal in real hardware, we make use of Information Content Landscape Analysis (ICLA)~\cite{perez-salinas2024analyzing} and find that the gradient does not vanish with increasing  circuit length.
This is  in agreement with the theoretical predictions in refs.~\cite{Pat2025beyondunitalnoise,meleNonunitalNoise2024}. 
Our findings indicate  that under real conditions NIBP will not be present, in apparent contradiction to  previous understanding~\cite{WangNIBP2021,Larocca2025ReviewBP}.
Nonetheless, the appearance of a finite gradient does not necessarily imply trainability and accuracy in the optimization of the VQA.
Finally, combining the analytical lower-bound for the absence of NIPB from ref.~\cite{Pat2025beyondunitalnoise} and the circuit runtime at which the gradient signal levels off allows us to estimate an effective \(\effTone\) of the hardware computations.
We find that \(\effTone\) is significantly smaller than the mean \(\Tone\) of the quantum device from the calibration data.
This  result is an indication that standard benchmark and calibration metrics based on mean properties of the hardware should not be fully trusted to estimate the computational capabilities of a quantum device.
As such, we envision the use of the methodology presented here to be used as yet another metric to more accurately estimate the actual \(\Tone\) for a particular computation.
To the best of our knowledge this is the first demonstration of the absence of NIBP and ICLA in actual quantum processing units.

The manuscript is organized as follows:
In the first part of~\Cref{sec:results}, we provide the necessary background for non-expert readers which include a description of variational quantum algorithms~\Cref{subsec:vqa}, information content landscape analysis~\Cref{subsec:icla}, a basic introduction to noise in quantum hardware and noise-induced barren plateaus~\Cref{subsec:noise}, and finally a general theory of barren plateaus~\Cref{subsec:bp}.
An expert reader might skip some or all of these subsections.
The second half of~\Cref{sec:results} is devoted to the experiment and the main results of this work.
In~\Cref{subsec:description_exp} we detail the experimental set-up, and the parameters of the IBM quantum processing units, followed by the presentation of the main figures in~\Cref{subsec:num_exp_absenceNIBP}, and the connection of the abscense of NIBP to an effective coherence time in~\Cref{subsec:TonEff}.
The manuscript continues with~\Cref{sec:discussion} where we discuss the implications of this study, lay down open research questions, and present some research directions for the future.
Finally, in~\Cref{sec:methods} we present further details of the experimental set up to ease its reproducibility, as well as additional calculation to support our main results.

\section{Results}\label{sec:results}
\subsection{Variational quantum algorithms}\label{subsec:vqa}
At the core of Variational Quantum Algorithms (VQA)~\cite{cerezo2020variational,bharti2022noisy,preskill2018quantum} lies a Parametrized Quantum Circuit (PQC) which generates a tunable quantum state through real parameters $\vtheta = (\vtheta_1, \dots, \vtheta_L)\in\mathbb{R}^m$,
\begin{equation}\label{eq:final-state}
    \rho(\vtheta)=U(\vtheta)\rho_0U(\vtheta)^\dagger,
\end{equation}
with an initial state $\rho_0 = \ket{\psi_0}\bra{\psi_0}$ and $U(\vtheta)$ the unitary operator representing the quantum circuit.
A common implementation of $U(\vtheta)$ is a $L$-layer variational circuit 
\begin{equation}\label{eq:PQC}
    U(\vtheta) = \prod_{l=1}^LU_l(\vtheta_l).
\end{equation}

The target of a VQA is to find the solution of a problem of interest encoded in a quantum observable \(\Obs\).
Let $\Obs$ be a $2^N\times2^N$ Hermitian matrix acting on the $N$-qubit Hilbert space.
Any such $\Obs$ can be expressed in the basis of $4^N$ Pauli operators \({\rm P}_i \in \{I, X, Y, Z\}^{\otimes N}\),
\begin{equation}
    \Obs = \sum_{i} c_i {\rm P}_i, \hspace{15pt} c_i \in \mathbb{R},
\end{equation}
which is straightforward to measure on a digital quantum computer.
Obtaining the cost function reduces to measuring each \({\rm P}_i\) w.r.t. to \(\rho(\vtheta)\) to reconstruct the expectation value of $\Obs$,
\begin{equation}\label{eq:exact-cost}
    C(\vtheta) = \trace[\Obs\rho(\vtheta)] = \sum_i c_i\langle {\rm P}_i\rangle = \sum_ic_i\trace[{\rm P}_i\rho(\vtheta)].
\end{equation}
In practice, one can only get statistical estimators of \({\rm P_i}\), $\E_R({\rm P}_i)$, following the Born rule.
The statistical estimator implies that \(R\) repetitions or copies of the quantum are needed and an approximation to the  expectation can be obtained 
\begin{align}\label{eq:samp-cost}
    C(\vtheta)&\approx\sampcost_R(\vtheta) = \sum_{i}c_i\E_R({\rm P}_i) \\
    & =
    \frac1R\sum_{z_r\sim p_{\vtheta}(z)}^R \sum_ic_i\trace[{\rm P}_i\ketbra{z_r}{z_r}]\, ,
\end{align} 
where \(p_{\vtheta}(z)=\trace\big(\rho(\vtheta)\ketbra{z}{z}\big)\) is the distribution of bitstrings $z\in\{0,1\}^N$ defined by the quantum state.
A classical optimization algorithm iteratively updates $\vtheta$ to minimize $\sampcost_R(\vtheta)$ until some convergence criteria is met.
The output of the algorithm is an approximation to the solution of the problem encoded in \(\Obs\).
Unfortunately, navigating even the noiseless quantum optimization landscape generated by this cost function is an arduous task due to its high multi-modality~\cite{bonet-monroig2023performance,deller2023quantum}.

\subsection{Information content landscape analysis}\label{subsec:icla}
BP are a property of the gradient \(\Grad\) of the variational quantum landscape.
Components of the gradient can be obtained by performing derivatives of the cost function along single directions
\begin{equation}
    \Grad \sim \frac{\partial C(\vtheta)}{\partial \theta_k}\,.
\end{equation}
Estimating the gradient on quantum hardware is typically done using analytic expressions, finite differencing schemes, parameter shift rules or cost function sampling~\cite{Arrasmith_2022, SchuldPSR2019, mariGradients2021}.

An alternative approach is to use data-driven methods to efficiently measure global properties of the variational quantum landscape.
Particularly useful for the analysis of BP is that of Information Content Landscape Analysis (ICLA)~\cite{perez-salinas2024analyzing,munoz2015ELA}.
In~\cite{perez-salinas2024analyzing}, the authors analytically prove that ICLA is an efficient and robust method to estimate the average norm of the gradient \(\Normgrad\), requiring only \(M=\bigO(m)\) independent sets of \(m\)-dimensional variational parameters \(\Theta = \{\vtheta_1, \dots, \vtheta_{M}\}\).
Then, the discrete cost function landscape \(\Omega = \{\sampcost_R(\vtheta_1), \dots, \sampcost_R(\vtheta_M)\}\) is obtained by measuring the cost functions for each parameter vector  with a quantum computer.

In~\Cref{alg:icla}, we provide a pseudo-code description of the ICLA algorithm adapted from~\cite{perez-salinas2024analyzing,munoz2015ELA}.
The output is an approximation to the average norm of the gradient
\begin{equation}\label{eq:normgrad}
    \Normgrad \simeq \epsM \sqrt{m},
\end{equation} 
where \(\epsM\) is a scaling parameter determined by ICLA signaling maximal information content of the cost function landscape.

\begin{algorithm2e}
\caption{Information Content Landscape Analysis (ICLA)}\label{alg:icla}
\textbf{Procedure:} ICLA(\(m\), \(F\),  \(C\))\;
\textbf{Input:} the number of parameters of the landscape \(m\), the factor of additional parameter sets \(F\), the cost function \(C\)\;
    Sample \(M(m) \in (F \cdot m)\) points of the parameter space \(\Theta = \{\vtheta_1, \dots, \vtheta_{M}\} \in [0, 2\pi)^m\)\;
    Measure \(C(\vtheta_i)\) on a quantum computer\;
    Store the sampled landscape \(\Omega = \{C(\vtheta_1, \dots, C(\vtheta_M)\}\)\;
    Generate a random walk \(W\) of \(S + 1 < M(m)\) steps over \(\Theta\)\;
    Compute the finite-size approximation of the gradient at each step \(i\) 
    \(\Delta C_{i} = \frac{C(\vtheta_{i + 1}) - C(\vtheta_{i})}{\norm{\vtheta_{i + 1} - \vtheta_{i}}}\)\;
    Create a sequence \(\phi(\epsilon)\) by mapping \(\Delta C_i\) onto a symbol in \(\{-, \odot, +\}\) with the rule
    \(\phi(\epsilon) = 
    \begin{cases} - & \text{ if } \Delta C_i < -\epsilon \\
    \odot & \text{ if } |\Delta C_i| \leq \epsilon \\
    + & \text{ if } \Delta C_i > \epsilon
    \end{cases}\)\;
    Compute the empirical IC at each \(\phi(\epsilon)\) from the empirical probabilities of consecutive symbols \(p_{ab}\): \newline
    \(H(\epsilon) = \sum_{a \neq b} -p_{ab} \log_6 p_{ab} \allowbreak p_{ab} =  \{p_{+-}, p_{-+}, p_{+\odot},p_{\odot +},p_{-\odot}, p_{\odot -}\}\)\;
    Repeat these steps for several values of \(\epsilon\)\;
    Find \(\epsM\) where \(H(\epsilon)\) is maximal: \(\epsM = \text{argmax}_{\epsilon} H(\epsilon)\)\;
    \textbf{Output:}  \(\Normgrad = \epsM \sqrt{m} \) (\Cref{eq:normgrad})
\end{algorithm2e}

\subsection{Noise in quantum hardware}\label{subsec:noise}
In this section, we provide a basic description of the most common noise sources in existing quantum hardware, with particular focus on the dominant effects in superconducting qubits.
For a more in-depth analysis we encourage the reader to follow~\cite{NielsenandChuang}.
A common description of noise in quantum hardware is \textit{depolarizing} noise.
In its simplest form, a single-qubit depolarizing model is described as
\begin{equation}\label{eq:depolarizing}
    \mathcal{E}_{\mathrm{dep}}(\rho) = (1-p)\rho + \frac{p}{3} \sum_{\substack{P \in \{ X,Y,Z\} \\ P\neq \mathbb{I}}} P \rho P
\end{equation}
describes the probability of a quantum state \(\rho\) to be randomly acted by a Pauli matrix with some probability \(p\).
Despite its simplicity, this unital model well captures the effects of noise when the quantum state is constantly monitored (e.g. quantum error correction).
However, in VQA one avoids intervention with the quantum state until measurement to (potentially) reduce the amount of errors.
As such, the depolarizing model is no longer a good description of the noise effects.

In near-term quantum algorithms, it is necessary to build noise models that more faithfully capture the processes at the hardware level.
Furthermore, they allow us to build benchmarking metrics to compare device performance across different qubit platform.
Rather popular such  metrics are the qubit coherence times \Tone and $T_\phi$,
characterizing amplitude damping and pure dephasing, respectively. 
The \(\Tone\) coherence time parametrizes the decay of a qubit from the exited state $\ket 1$ to the ground state $\ket 0$ with a probability, $p_A(t)$, after some time \(t\):
\begin{equation}\label{eq:decays}
    p_{A}(t) = 1 - e^{-\frac{t}{\Tone}}\,.
\end{equation}
The time evolution of a single qubit density matrix \(\rho(t)\) under amplitude damping is described by 
\begin{equation}
    \rho_{A}(t) =\begin{pmatrix} 
            \rho_{00}+(1-e^{-\frac{t}{T_1}})\rho_{11} & e^{-\frac{t}{2T_1}}\rho_{10}\\
            e^{-\frac{t}{2T_1}}\rho_{01} & e^{-\frac{t}{T_1}}\rho_{11}
            \end{pmatrix}.
\end{equation}
A more realistic noise model additionally includes a dephasing noise channel characterized by $T_2$ 
\begin{equation}
\label{eq:T1T2noise}
    \rho_{A+D}(t) =\begin{pmatrix} 
            \rho_{00}+(1-e^{-\frac{t}{T_1 }})\rho_{11} & e^{-\frac{t}{ T_2}}\rho_{10}\\
            e^{-\frac{t}{T_2 }}\rho_{01} & e^{-\frac{t}{T_1 } } \rho_{11}
            \end{pmatrix},
\end{equation}
where \(T_2=\left(\tfrac1{2\Tone}+\tfrac1{T_\phi}\right)^{-1}\). The former represents a non-unital noise channel, whereas pure $T_\phi$ dephasing is unital.

These equations follow from a Kraus operator description of the quantum processes which we avoid for simplicity in this manuscript, but can be found in~\cite{NielsenandChuang}.
For sufficiently long times \(t \gg T_{i}\), the previous two equations describe the decay of the density matrix into a pure state  $\rho = \ket{0}\!\bra{0}$ or a mixed state $\rho = \rho_{00}\ket{0}\!\bra{0} + \rho_{11}\ket{1}\!\bra{1}$.

While coherence times are of great importance to understand the limits of near-term quantum hardware there exist other relevant sources of error.
First, execution of single- and two-qubit gates have  small errors which accumulate and degrade  the quantum computation.
Measurement errors which fail to correctly classify the final state of qubit.
Device cross-talk between neighboring qubits.
Qubit frequency drifts, two-level systems, and cosmic ray bursting just to mention some of them.
More comprehensive details about noise sources in superconducting qubits can be found in~\cite{IntroCQEDVoolDevoret,ReviewCQEDBlais2021}.

\subsection{Barren Plateaus}\label{subsec:bp}
Since the advent of VQA, significant effort has been devoted to finding their trainability limits and its potential connection to classical simulability~\cite{CerezoAbscenceBP2025}.
A central obstacle to trainability is the onset of BP defined as the exponential decay of the variance of the gradient with system size \(N\)~\cite{mcclean2018barren,cerezo2021higher-order},
\begin{equation}\label{eq:BPs}
    \Var_{\vtheta}[\partial C(\vtheta)]\in\mathcal{O}\big(\frac{1}{b^N}\big)
\end{equation} 
for some $b>1$.
Four aspects of a VQA have been identified as sources of BP:
\begin{inparaenum}[(i)]
\item Expressivity of the circuit~\cite{Fontana2024BpLieTheory,holmesExpressivityBP2022},
\item Entanglement of the accessible quantum states~\cite{Ragone2024LieAlgebraBP, thanasilp2023inputstateQML},
\item Cost function or observable locality~\cite{Ragone2024LieAlgebraBP},
\item Noise in quantum hardware~\cite{WangNIBP2021}.
\end{inparaenum}

Among all these sources, NIBP are of particular interest as they arise independently of the system size \(N\), observable, or entanglement and solely depend on the circuit length or runtime.
An in-depth theoretical analysis of NIBP is done in~\cite{WangNIBP2021} where the authors assume a single-qubit Pauli error model, a simplified version of the depolarizing noise model described in~\Cref{eq:depolarizing}.
The consequence of this assumption is significant; given a sufficiently large circuit (e.g.\ polynomially growing with \(N\)), the final state will tend exponentially towards the maximally mixed sate.
In turn, this implies that any measurement over the final state will draw bitstrings from a uniform distribution, and expectation values will always be the same, regardless of the actual circuit parameters.
Estimating the gradient over the cost function  will then always lead to a value arbitrarily close to zero, and thus BP.

As already stated in~\Cref{subsec:noise}, the assumption of a depolarizing noise model is a  simplification of the  actual processes occurring in quantum hardware.
More realistic noise models that include non-unital channels have been proven to not produce NIBP~\cite{Pat2025beyondunitalnoise,feffermanNonUnitalNoisePRXQ2024,meleNonunitalNoise2024}, or alternatively a finite gradient.
Our experiments corroborate this prediction, and raise the question of whether NIBP can actually be observed experimentally.
The reason for this is the fact that non-unital channels are unavoidable in quantum hardware, and following the analysis in~\cite{Pat2025beyondunitalnoise} there will always exist a circuit length for which NIBP will not appear.

\subsection{Experimental setup on IBM quantum hardware}\label{subsec:description_exp}
To experimentally probe the impact of amplitude damping on  NIBP, we study the nearest neighbor Ising chain
\begin{equation}\label{eq:ising} 
    {\rm H}_C= \sum_{i=0}^{N-2}\frac{J_{i,i+1}}{2}Z_iZ_{i+1}+\sum_{i=0}^{N-1}\frac{h_i}{2}Z_i 
\end{equation}
as the model Hamiltonian for the cost function.  
The parameters $J_{i,i+1}$ and $h_i$ are drawn randomly from a set of fixed values 
\(J_{i,i+1}\sim\mathcal{U}(\{\pm 2, \pm 1.2, \pm 0.8, \pm 0.4\}) \)  and 
\( h_i\sim\mathcal{U}(\{0.8, \pm 0.4, \pm 0.24, \pm 0.16, -0.08\})\).
We chose this Hamiltonian  for two reasons. For one, it is $2$-local so cost function dependent BP do not arise.
For another, the coupling topology  is compatible with the IBM quantum hardware layout, so there is no overhead in terms of two-qubit gates to implement this model.  

We chose a typical Quantum Approximate Optimization Algorithm (QAOA)~\cite{farhi2014quantum,blekos2024QAOA_Review,bharti2022noisy} ansatz for the PQC which consists  of 
$L$ alternating cost and mixing layers generated by ${\rm H}_C$ and ${\rm H}_M=\sum_{i=0}^{N-1}X_i$, i.e.\
\begin{equation}\label{eq:ising-circuits}   
    U(\vtheta)= \prod_{l=1}^{L}e^{-i\theta^{(1)}_l {\rm H}_M}  e^{-i\theta^{(2)}_l {\rm H}_C},
\end{equation}
where we have split $\vtheta$ into mixing parameters, $\vec{\theta}^{(1)} = (\theta^{(1)}_1,\dots,\theta^{(1)}_L)$, and phase parameters, $\vec{\theta}^{(2)} = (\theta^{(2)}_1,\dots,\theta^{(2)}_L)$.
For this ansatz the number of layers directly determines the dimensionality of the parameter vector as $m=2L$.

Starting from the equal superposition of all computational basis states as initial state, $\rho_0 = \ket{...+...}\!\bra{...+...}$,
the quantum circuit produces the output state \(\rho(\vtheta) = U(\vtheta)\rho_0U(\vtheta)^\dagger\).

For a given set of parameters $\vtheta$, we measure all qubits in the computational basis $R$ times which produces a set of bitstrings \(z\in\{0,1\}^N\). Each bitstring is drawn from the distribution defined by the quantum state,
\(z\sim  p_{\vtheta}(z)=\trace\big(\rho(\vtheta)\ketbra{z}{z}\big)\).
Because \Cref{eq:ising} contains only commuting $Z_i$ and $Z_iZ_{i+1}$ terms, a single measurement setting is enough to estimate all contributions to the energy. The energy contribution of a single shot $z^{(r)}$ is given by
\begin{equation}
\label{eq:costSingle}
    E(z^{(r)}) = \sum_{i=0}^{N-2}\frac{J_{i, i+1}}{2}s^{(r)}_is^{(r)}_{i+1} + \sum_{i=0}^{N-1}\frac{h_i}{2}s^{(r)}_i,
\end{equation}
where $s_i^{(r)} = 2 z_i^{(r)}-1\in\{-1, 1\}$ are the shifted bitstring values.
All $R$ of these single shot energies are then averaged up to give an experimental estimate for the cost function value for a given parameter set of the circuit.

We transpile logical circuits given by \Cref{eq:ising-circuits} with default setting on the IBM qiskit transpiler.
Detail on the gate statistics and circuit depths, as well as representative examples of the circuit used in this study are given in \Cref{subsec:circuits_hardware}.

We used ICLA to estimate the norm of the gradient of the cost landscape for $N\in\{8, 20, 45, 65, 102\}$ qubits and  number of layers up to $L=120$. 
The landscape was created using $M=\min(10m,200)$ parameter vectors, where $m=2L$ is determined by the number of layers of the quantum circuits. 
We truncated the size of the landscape to $M=200$ for large systems as we observed this to be sufficient. 
See \Cref{sec:robustICLA} for more details on this. 
This observation makes the ICLA even more efficient as it substantially reduces the number of required circuit evaluations for large circuit depth $L$. 

We estimate the cost function value by using $R=16384$ measurement shots for each parameter set. 
We ensured that the shot-noise error was always at least a factor of two  below the observed gradient signal. 
This forced us to perform some few experiments with $R=32768$ and $R=65536$ measurement shots.
We refer the reader to \Cref{subsec:costFunctions} for more details on shot-noise floor and how cost function values are calculated.

Typical error mitigation techniques~\cite{errorMitigationReview2023} like Pauli-twirling, dynamical decoupling and zero-noise extrapolation are not effective in mitigating non-unital amplitude damping errors. 
We explicitly confirmed this by performing dynamical decoupling and Pauli twirling with 32 and 64 randomizations and $R=4095$ shot per randomization for $N=68$ qubits and with up to $L=30$ layers. 
The resulting gradient signals were slightly lower than without error mitigation, as it can be expected when some noise sources are reduced and  therefore, the noise in the cost function landscape is also reduced. 
But the qualitative flattening behavior of the gradient signal for long runtime (large layers) was observed as well.

We ran experiments on IBM $127$-qubit  Falcon processors \texttt{ibm\_brisbane}, \texttt{ibm\_kyiv}, and \texttt{ibm\_sherbrooke}, as well as the 156 qubit Heron processor \texttt{ibm\_fez}.
For each experimental run we recorded the assignment of logical to physical qubits and the backend calibration data, in particular \(\Tone\), \(\Ttwo\), ECR-gate error, and readout error.
More details on the hardware calibration data can be found in Appendix~\ref{subsec:circuits_hardware}, in particular Table.~\ref{tab:calib_all}.

\subsection{Numerical and experimental evidence of the absence of NIBP}\label{subsec:num_exp_absenceNIBP}
\begin{figure}
    \centering
    \includegraphics[width=\columnwidth]{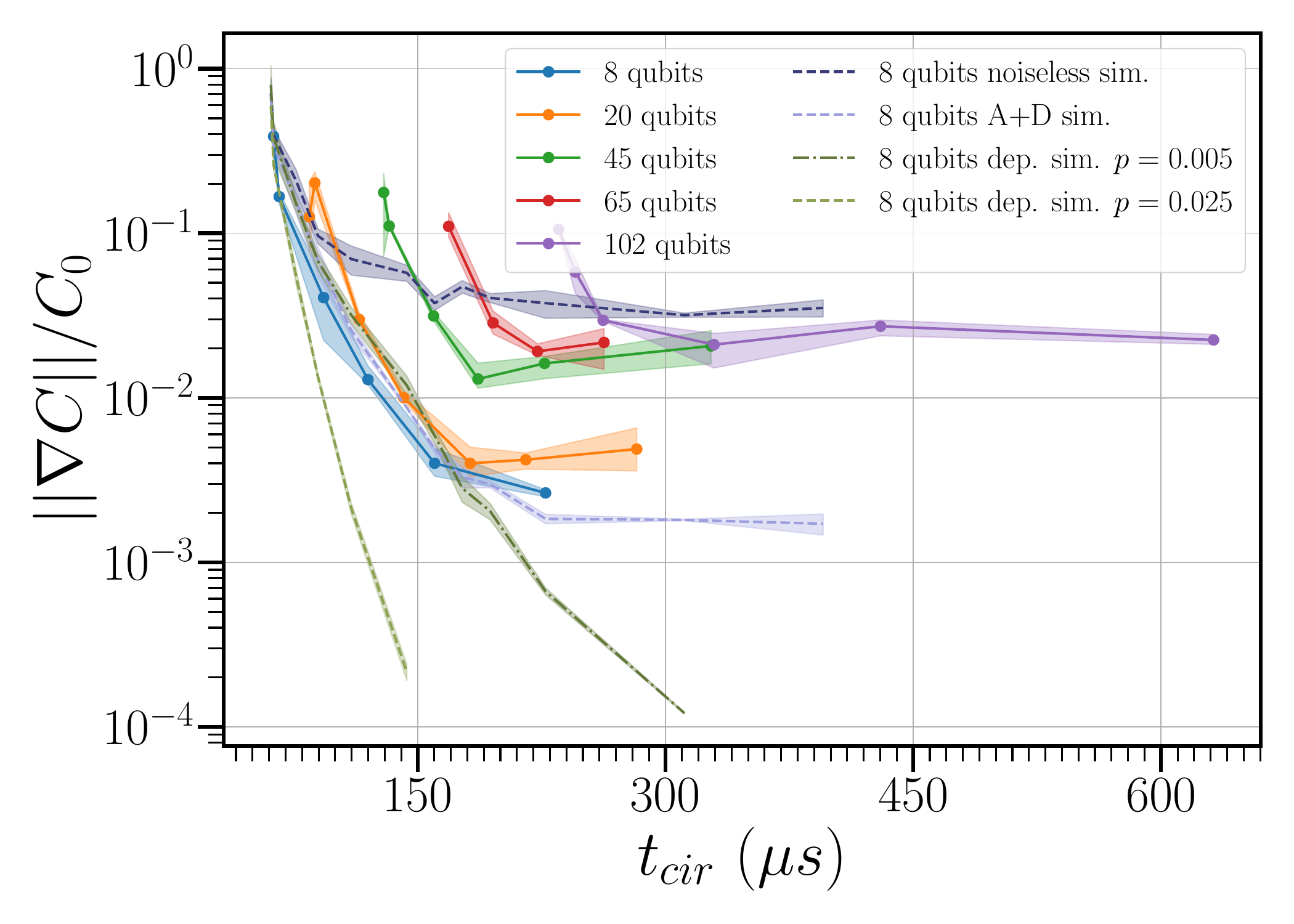}
    \caption{Average norm of the gradient (\(\Normgrad\)) versus circuit runtime (\(\cirtimes\)).
    Here, \(\cirtimes\) only accounts for the time of the gate implementation and measurement, but we removed the reset time between shots.  
    Lines with markers show the data obtained from actual quantum hardware.
    Each color indicates the qubit size of the experiment.
    Lines without markers depict the 8-qubit density matrix simulations, where 'dep.' refers to depolarizing noise with strength $p$ of Eq.~\eqref{eq:depolarizing} and "A+D" denotes the noise model of Eq.~\eqref{eq:T1T2noise}. 
    While depolarizing noise simulations show clear signal of NIPB, the gradient signal flattens at a constant value for large circuit runtimes for all hardware experiments and amplitude damping simulations.
    }
    \label{fig:avg_grad_circuit_times_noNIBP}
\end{figure}

In this section we present the main results of our experiment using IBM quantum hardware and classical simulations. 
In~\Cref{fig:avg_grad_circuit_times_noNIBP} we depict the (estimated)  average norm of the gradient \(\Normgrad\).
We introduced the normalization parameter $C_0=\sqrt{\sum_i(J_{i,i+1}^2+h_i^2)}\sim\sqrt{N}$ to account for the  scaling of the Hamiltonian with the number of qubits $N$.
As a baseline comparison to the experimental data we include classical  density matrix simulations for $N=8$ qubits.

We start by discussing the results of the classical simulations (dashed lines).
For the simulated circuits we estimate the hardware runtime $\cirtimes$ by taking their depth and translating it using the  average time needed for a given circuit depth from the hardware runs (see \Cref{subsec:circuits_hardware} for more details). 
The  noiseless simulation (dark purple) does not show any sign of BP indicated by \(\Normgrad\) initially decreasing from \(0.5\), but then leveling off at \(0.05\) for runtimes above $\sim 150\mu s$.
This is the expected behavior in our experiment based on our choice of a local observable to avoid cost-function dependent BP~\cite{cerezo2021cost}.
Conversely, NIBP arise in the noisy simulations employing a single-qubit depolarizing noise model of Eq.~\eqref{eq:depolarizing}.
A clear exponential decay in \(\Normgrad\) going from \(1\) to \(10^{-4}\) is observed in a very short runtime, \(\cirtimes \lesssim 150 \mus\), for single-qubit depolarizing probability, \(p=0.025\) (light green).
The effect is less severe for a milder depolarizing probability, \(p=0.005\), where it takes approximately twice as long to reach the same value of \(\Normgrad\) (dark green).
In any case, we conclude that the depolarizing noise model inevitably leads to NIBP as extensively analyzed in~\cite{Larocca2025ReviewBP}.
We also simulate the experiment under a more realistic noise model that includes  \(\Tone\) and \(\Ttwo\) coherence times as described in Eq.~\eqref{eq:T1T2noise} (light blue).
We used randomized coherence times drawn from normal distributions $T_1\sim\mathcal{N}(\mu=244\mu s,\sigma=74\mu s)$ and $T_2\sim\mathcal{N}(\mu=159\mu s,\sigma=93\mu s)$  for this noise model, which are consistent with the coherence times from the hardware experiments.
At short circuit times, up to \(\cirtimes \lesssim 170 \mus\), both the realistic noise  and the weak depolarizing noise models have an almost identical decay, starting at \(\Normgrad \approx 0.5\) and exponentially decreasing  to \(\Normgrad\approx0.005\).
At \(\cirtimes \gtrsim 170 \mus\) to \(\cirtimes \approx 230 \mus\) the trend of \(\Normgrad\) in the realistic noise model changes drastically, and at \(\cirtimes\gtrsim 230 \mus\) onwards, it fully flattens to \(\Normgrad \approx 0.002\). These simulations confirm that non-unital noise in the form of amplitude damping prevents the appearance of  NIBP for runtimes on the order of the decoherence time, \(\cirtimes \sim \Tone\).  

In order to give some physical intuition for the vanishing NIBP, a spectral analysis of the simulated 8-qubit density matrices~\Cref{subsec:density-matrices} was carried out. 
The spectrum of the depolarizing noise model indicated a fully mixed state for large circuit depth (long runtimes). 
On the other hand, the spectrum of the amplitude damping noise model was confined to a noise-induced limit set as described in ref.~\cite{Pat2025beyondunitalnoise}.
The noise-induced limit set is characterized by a density matrix where the spectral weight is distributed over a finite subset of states. 
Because the noise-induced limit set depends on the variational parameters of the quantum circuit, the corresponding cost function values also depend on the variational parameters. 
Therefore, the noise-induced limit set leads to cost function values which are not exponentially concentrated around one value, therefore, a finite gradient as we observe.

Next we shift our attention to the experimental data in~\Cref{fig:avg_grad_circuit_times_noNIBP} (solid marked lines).
The number of qubits in our experiments range from $N=8$ to $N=102$ qubits with circuit times of \(\cirtimes = 50-630 \ \mus\).
We observe some general trends; first, all experimental data start at a similar landscape gradient, \(\Normgrad\approx0.2\), despite starting at a different \(\cirtimes\).
Second, in all the curves \(\Normgrad\) initially decays rapidly, with a sudden change in trend at some \(\cirtimes\) which differs by the qubit number.
Looking closely at the data, the $N=8$ qubit case (blue) shows a remarkable agreement to the realistic noise model (dashed light blue).
Both lines start at \(\cirtimes=60\mus\) with \(\Normgrad\approx0.5\) and continue their decay with the same trend until \(\cirtimes \gtrsim 160\mus\).
At this point both curves level off at around \(\cirtimes\gtrsim230\mus\), with \(\Normgrad\approx0.002\).
While we do not have further experimental data in the 8-qubit case, the qualitative agreement between the simulation and the experimental curve gives us confidence that this trend will continue.

Moving to the \(N=20\) experiment (orange), we observe that \(\Normgrad\) behaves qualitatively similar to the 8-qubit experiment, reaching the anticipated flattening  around \(\cirtimes\approx 180\mus\), where the gradient sets at \(\Normgrad\approx0.004\).
The non-uniform trend for very short circuits (first two points) is an artifact of ICLA for very small numbers of sampling points of the cost landscape. 
Identically, for the \(N=45\) (green) experiment we observe a gradient decrease, followed by a constant gradient regime with \(\Normgrad\approx0.01\) at \(\cirtimes\gtrsim 180\mus\). 

The largest experiments were done with \(N=65\) (red) and \(N=102\) (purple) qubits, also show the same qualitative behavior.
After a decrease the gradient levels off around $\cirtimes\gtrsim 230\mu s $ for $N=65$, and $\cirtimes\gtrsim 270\mu s$ for $N=102$,  with value \(\Normgrad\approx 0.025\). 
For the largest circuits considered here $N=102$, this trend is observed up to very long times of \(\cirtimes=630\mus\), clearly marking the absence of NIBP.

The results for \(N=20\) and \(N=45\) in Fig.~\ref{fig:avg_grad_circuit_times_noNIBP} show a small upward trend for long runtimes, while no such increase is observed for $N=65$ and $N=102$.
We can link the appearance of this upward trend to  different calibrations and, in particular, different physical qubits being used for different circuit lengths. 
For example, the last three points for $N=20$  and $N=45$ use different physical qubits,  where the mean $\Tone$ varies between $\ToneMean\approx 245\mu s$ and $\ToneMean \approx265\mu s$ for $N=20$,  and between $\ToneMean\approx220\mu s$ and $\ToneMean\approx242\mu s$ for $N=45$.
In contrast, the last points for the cases of $N=65$ and $N=102$ used the same physical qubits with only slightly different calibrations where mean $\Tone$ varies very little (between $\ToneMean\approx247\mu s$ and $\ToneMean\approx249\mu s$ for $N=65$, and between $\ToneMean\approx236\mu s$ and $\ToneMean\approx238\mu s$ for $N=102$).
Such statistical variations in the coherence times lead to additional noise in the cost function values. 
For the ICLA this implies more structure in the landscape which directly leads to an increase of the gradient.  

The results presented here robustly show the absence of NIBP on IBM quantum hardware \texttt{ibm\_brisbane} for long circuit runtimes with circuits involving between $N=8$ and $N=102$ qubits.    
The gradient signal levels off to a constant value  above a characteristic circuit runtime which is determined by the amplitude damping coherence time \Tone  of the qubits. 
For circuit runtimes longer than this scale, the structure of the cost function landscape stays the same which is consistent with a noise-induced limit set~\cite{Pat2025beyondunitalnoise}.
The circuit depth which corresponds to the flattening of the gradient defines an effective circuit depth $L_\text{eff}$ as described in ref.~\cite{meleNonunitalNoise2024}.  
Circuits longer than this depth behave as if they have the shorter effective depth where only the variational parameters of the last $L_\text{eff}$ layers influence the output state significantly.
As already discussed in ref.~\cite{meleNonunitalNoise2024}, the effective depth is also limiting the expressiveness of  the quantum circuit and makes them ineffective for variational quantum optimization problems.
This is also in accord with the implications of a noise-induced limit set, where also for very deep circuits only a small set of different cost functions values can be attained. 

The findings of these works also have implications for the interpretation  of the results of other recent works~\cite{barreraBenchmarking2025,barreraLR-QAOA2025,pelofske2025evaluating}, where  similar circuits are employed to solve optimization problems on actual quantum hardware.   
Crucially, the authors do not consider amplitude damping, and always assume the the fully mixed state is reached for large circuits. 
However, we suggest that it is important to consider amplitude damping and the suppression of NIBP, especially when analyzing the performance degradation for large circuit depth. 
In particular when comparing different hardware platforms  with very different structure for the gate and coherence times.
For example, it would be very interesting to reanalyze the much weaker decay of the success probabilities with circuit depth on trapped ion devices (see, for example, Fig.~6(b) in ref.~\cite{barreraLR-QAOA2025} and Fig.~4(a) in \cite{barreraBenchmarking2025}) in light of the relatively much longer $\Tone$ coherence times on these platforms as compared to IBM superconducting devices.  

Interestingly, another recent work \cite{schumannNoise2024}  observed a very similar behavior as we do here. 
The results showed that the small circuit signal (purity in their case) decreases significantly with increasing amplitude damping noise, but then also  flattens out for large circuits and  for large noise strength (see Fig.~4 of ref.~\cite{schumannNoise2024}).
While the authors acknowledge that the fully mixed states is not a fixed point under amplitude damping, and that therefore the purities flatten out, they still conclude that amplitude damping incurs a NIBP.
However, we would  argue instead that flattening  signals the suppression of a  NIBP, and that they actually observed the residual variance of the noise-induced limit set.  

\subsection{Effective \(\Tone\) from the absence of NIBP}\label{subsec:TonEff}
So far, we have been able to validate the theoretical predictions from ref.~\cite{Pat2025beyondunitalnoise}, that  under sufficiently large non-unital noise the cost function values do not concentrate, i.e.\ NIBP are absent.
We now turn to the topic of relating the qubit coherence times $\Tone$ and decay probabilities  to the flatting time using the insights from Ref.~\cite{Pat2025beyondunitalnoise}.
The authors analytically predict the absence of NIBP when the probability of non-unital noise, i.e.\ the decay probability of a single qubit into its ground state, is \(p_{{\rm non-unital}} = \frac{3}{4}\). 
We assume that the onset of the regime where  non-unital noise dominates is signaled by the flattening of the gradient signal.
Therefore, we  conclude that the amplitude damping probability has reached  \( p_{{\rm non-unital}}\) at that runtime where the gradient flattens, i.e.\ when $\cirtimes=\cirtimes^\text{flat}$. 
We can use~\Cref{eq:decays} to relate the circuit runtime to the decay probability and estimate an effective decoherence time,
\begin{equation}\label{eq:efft1}
    \effTone = \frac{- \cirtimes^\text{Flat} }{\ln[1 - p_{{\rm non-unital}}]}  
    \approx 0.72\, \cirtimes^\text{Flat} .
\end{equation}
This effective coherence time is the relevant timescale for estimating a single qubit’s decay probability and assessing how amplitude damping flattens the gradient, i.e., suppresses NIBP.
\begin{table}[ht]
\centering
\caption{Mean and effective $\Tone$ along with the qubit percentages below these values, and the flattening times $\cirtimes^{\text{flat}}$ extracted from the data.  }
\label{tab:effT1}
\begin{tabular}{|c|c|c|c|c|c|}
\hline
 $N$& \textbf{$\langle\Tone\rangle [\mus]$} & \textbf{\% qubits}  & \textbf{$\cirtimes^\text{flat} [\mus]$} & \textbf{$\effTone [\mus]$} & \textbf{\% qubits} \\
\hline
8   & $216  \pm 39$  & 32 & $230\pm20$  & $166\pm 33$ & 13 \\
\hline
20  & $256\pm65 $ & 52 & $180\pm20$ &$130\pm36$  & 2.8 \\
\hline
45  & $240 \pm76$ & 50 & $185\pm15$ & $134\pm44$ & 7.4 \\
\hline
65  & $247\pm 77 $  & 56 & $220\pm15$ & $166\pm51$ & 11 \\
\hline
102 & $236 \pm68$ & 51 & $260\pm30$ & $188\pm58$ & 24 \\
\hline
\end{tabular}
\end{table}
We compare the effective coherence time $\effTone$ with the actual coherence times $\Tone$ of the circuits used for the experiments.
\Cref{tab:effT1} shows the mean coherence times of the experiments on \texttt{ibm\_brisbane} device for various qubits, the flattening times estimated from our results as well as the corresponding estimated effective coherence times \effTone. 
We also report the fraction of qubits which have a coherence time below the mean and the effective coherence times.
It needs to be noted that the extracted flattening times \( \cirtimes^\text{flat}\)  have a large error associated with them due to the rather coarse sampling of circuit times, see~\Cref{fig:avg_grad_circuit_times_noNIBP}.

Naively one would expect that the onset of amplitude damping effects is determined  by the mean coherence times of the used qubits.
However, we observe that the effective coherence times \effTone are substantial smaller than the mean $\ToneMean$, which implies that effects of amplitude damping are observable much earlier.  

\begin{figure}
    \centering
    \includegraphics[width=\columnwidth]{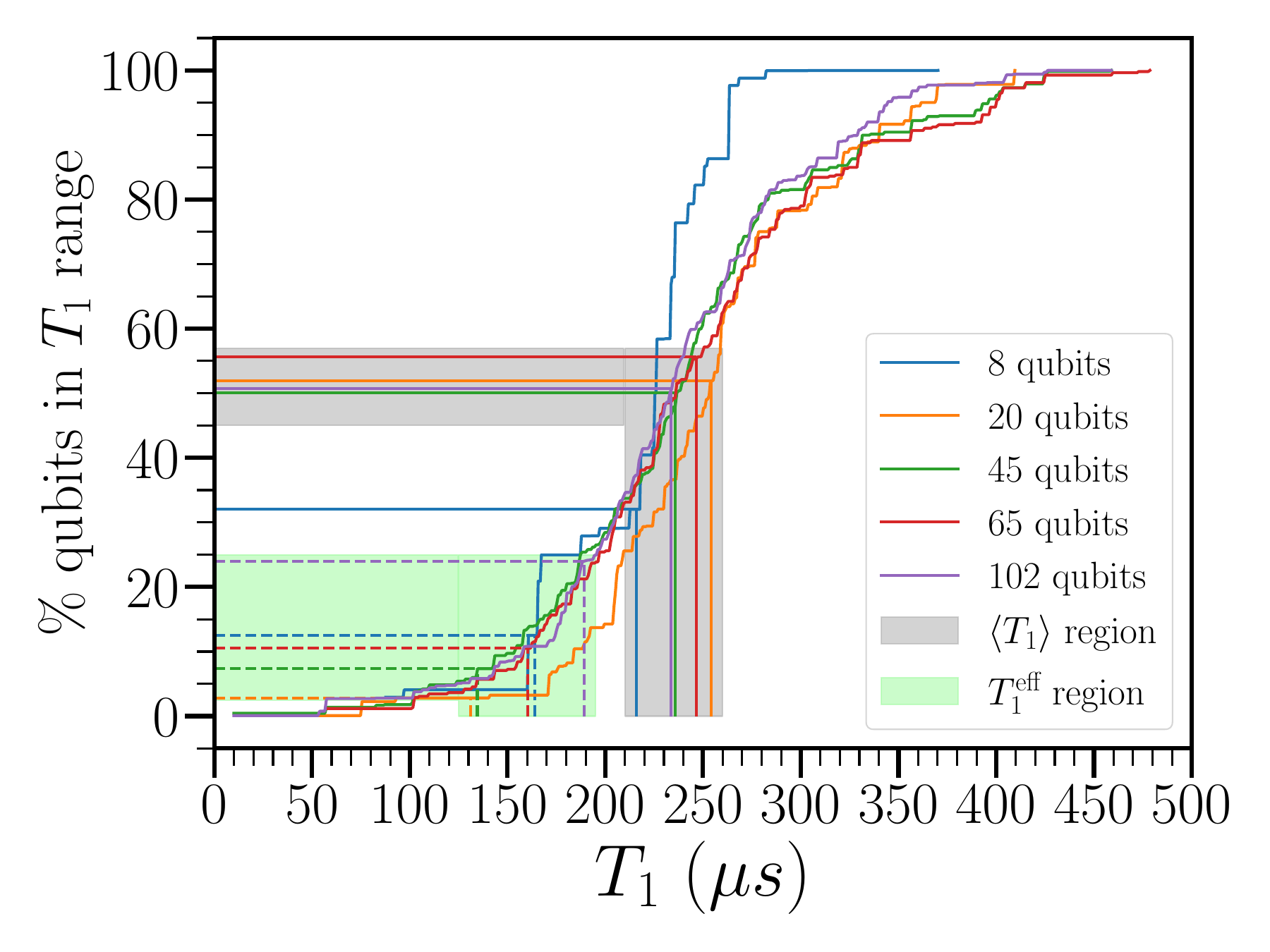}
    \caption{\(\Tone\) cumulative distribution function versus \% of qubits contained in the \(\Tone\)-range. Qubit numbers are marked with colored lines. For every circuit size, the vertical and horizontal lines indicate the mean (solid), and effective (dashed) \(\Tone\). The mean \(\ToneMean\) region is denoted with the shaded gray area. The effective \(\effTone\) region is denoted with the green shaded area. 
    This result  suggests that the effective $\effTone$ is determined by the $~\sim20\%$ of the qubits with the shortest coherence times.
    }
    \label{fig:t1_cdf_vs_qubits}
\end{figure}

In~\Cref{fig:t1_cdf_vs_qubits} we show the cumulative distribution function (CDF) of the \(\Tone\) values from the \texttt{ibm\_brisbane} device  against  the percentage of qubits with a  \(\Tone\) up to the indicated  value.
The values of the mean and effective \(\Tone\) as well as the qubit percentage within this range are indicated in the figure.
The solid (dashed) horizontal and vertical lines indicate the mean (effective) \(\Tone\) for each circuit size.
The colored areas indicate the regions of \(\Tone\) and qubit percentage for the mean (gray) and effective (green) coherence times.
As expected, the mean \(\Tone\) roughly equates to \(50\%\) of the qubits.
On the other hand,  only about 3\%-25\% of the qubits have a coherence time equal or  lower than  the respective effective $\effTone$.   
This result clearly indicates that the expected circuit times at which \(\Normgrad\) should level off are much shorter than they are naively expected from the mean coherence times $\ToneMean$.
Put  differently, we could conclude that it is the $\sim 20\%$ of the qubits with the lowest coherence times which determine the flattening of the gradient signal.

We validated these insights by running more experiments with different setups.
First, we ran the same circuits on the different hardware platforms.
The \texttt{ibm\_kyiv} and \texttt{ibm\_sherbrooke} platform are the same Falcon architecture as \texttt{ibm\_brisbane} and only differ in the distribution of coherence times and errors. 
In contrast,  the \texttt{ibm\_fez} platform is of the Heron architecture where different gates with much shorter gate times are implemented. 
However, the coherence times of \texttt{ibm\_fez} are also substantially  shorter as can be seen in \Cref{subsec:circuits_hardware}.
Second, we ran the circuits more efficient gate sequence as described in \Cref{subsec:circuits_hardware}, which leads to much shorter runtime on  \texttt{ibm\_brisbane} for $N=45$.
And lastly, we also implemented a slightly modified Hamiltonian  where the coupling topology was exactly that of the heavy hex hardware coupling.

\begin{figure}
    \centering
    \includegraphics[width=\columnwidth]{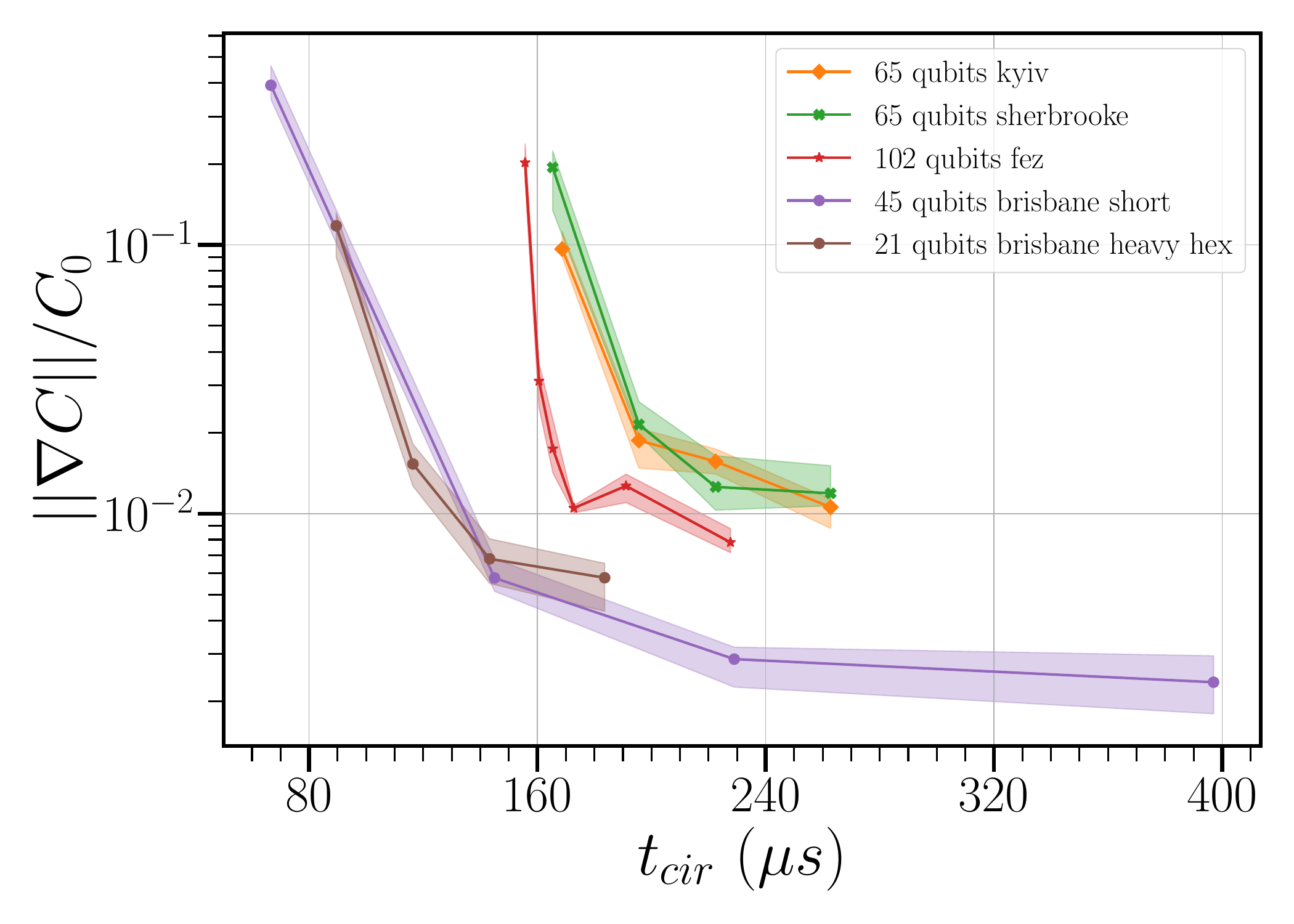}
    \caption{Gradient, \(\Normgrad\), as a function of circuit times \(\cirtimes\) for various setups. 
    The data labeled 'short' refers to the different compilation of the same logical circuit (see \Cref{subsec:circuits_hardware}), while 'heavy hex' indicates that a different Hamiltonian with same coupling topology as the hardware was used.
    These results show that the flattening phenomenon is consistent across different  hardware and setups.
    }
    \label{fig:otherGrad}
\end{figure}

The resulting gradients for those cases can be observed in \Cref{fig:otherGrad}.
Regardless of the hardware platform, circuit type and Hamiltonian, the qualitative behavior is  the same as observed before. 
After an initial rapid decrease, the gradient signal flattens out  and approaches a constant value. 
The remaining variations, in particular for the case of \texttt{ibm\_fez}, can again be related to different calibrations with significantly changed coherence times for the used qubits.  

\begin{figure}
    \centering
    \includegraphics[width=\columnwidth]{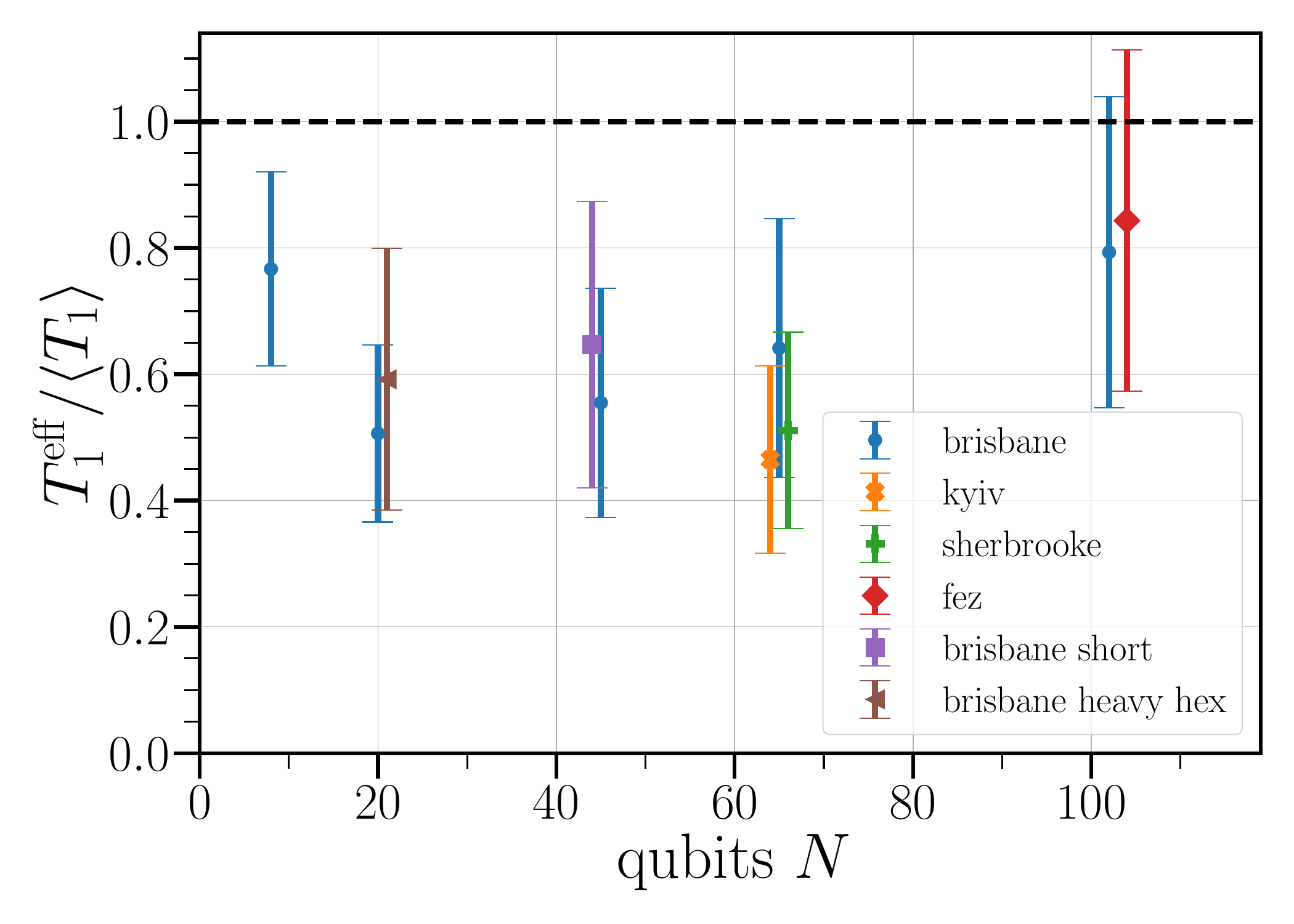}
    \caption{\(\effTone\) relative to \(\ToneMean\) as a function of the number of qubits and for various different experimental setups. 
    Each color represents a different \texttt{ibm} device. 
    'short' (purple) and 'heavy hex' (brown) represent different compilation and a different topology for the Hamiltonian, respectively.
    The effective coherence times $\effTone$ are significantly smaller than mean coherence times $\ToneMean$ of all qubits in the circuits.
    }
    \label{fig:teffall}
\end{figure}

Finally, we show the effective coherence times determined from all experiments relative to the mean $\Tone$ as function of number of qubits $N$ in \Cref{fig:teffall}.
The observed values for $\effTone$ are all well below \ToneMean, and all lie within a range of $0.4\lesssim \effTone/\ToneMean\lesssim 0.9$.
This reconfirms the finding that amplitude damping dominates gradient signal once the runtime of the circuit approaches the effective coherence time which is determined, but slightly below the mean coherence times of the qubits of the circuits.

The determination of the effective coherence times relies on the lower bound for the decay probability of   \(p_{{\rm non-unital}} = \frac{3}{4}\) as derived in ref.~\cite{Pat2025beyondunitalnoise}.
If, however, a smaller threshold probability for the dominance of amplitude damping noise is assumed, the resulting effective $\effTone$ will be larger. 
Based on our experimental data, the effective and the mean coherence times would coincide for a threshold probability in the range \(p_{{\rm non-unital}} \sim 0.45-0.65 \). 
Therefore, finding more accurate estimations of the theoretical threshold probability is important for an accurate estimation of  \effTone.

\section{Discussion}\label{sec:discussion}
In summary, we have experimentally demonstrated the absence of Noise-Induced Barren Plateaus (NIBP) due to the effects of non-unital noise as theoretically predicted in refs.~\cite{Pat2025beyondunitalnoise,meleNonunitalNoise2024}.
We show that the gradient of the variational quantum landscape does not vanish  as the circuit runtime increases as it would be expected by the Barren Plateaus (BP) theory.
Instead,  the gradients approach a constant value for long circuit runtimes, i.e.\ circuit depth.

One of the contributions of this work is the use of Information Content Landscape Analysis (ICLA) to efficiently estimate the average norm of the gradient of the variational landscape in real quantum hardware.
Through ICLA, we compare noisy classical simulations against the same computation in real hardware, showing a remarkable qualitative agreement that confirms the absence of NIBP under non-unital noise.
Relying upon this comparison, we extend our experiment to larger qubit systems with longer circuit runtimes, different hardware platforms, and modified coupling topologies in the Hamiltonian.
All of the experiments robustly show the same qualitative behavior, where the average norm of the gradient tends to a constant value for long circuit runtimes.

Our analysis identifies the single qubit coherence times $\Tone$  as the main driver for the non-unital noise channel. 
Based on the theoretical prediction of ref.~\cite{Pat2025beyondunitalnoise} 
we derive an effective coherence time $\effTone$ which determines the characteristic time scale for the dominance of non-unital noise. 
The effective coherence times are directly related to the experimentally observed flattening times for the gradient. 
Comparing the extracted coherence times to calibration data, we find that  $\effTone$ is significantly smaller than the mean coherence time of the qubits, $\ToneMean$, used in the computation. 
In turn, this implies  that the circuit runtimes where non-unital amplitude damping is not affecting the results strongly are much shorter than one would expect from mean coherence times of the qubits.

Relating the observed $\effTone$ to the distribution of the coherence times on the real hardware, we show that the worst $\lesssim20\%$ of  qubits with the shortest  \(\Tone\) determine the effective  \(\effTone\) and thus the onset of the flattening of the gradient signal.
This suggests that average quantum device data can not be reliably used to estimate the impact of errors in a near-term quantum computation, but instead the actual distribution needs to be considered. 

Our work focused on the suppression of NIBP due to non-unital amplitude damping noise. 
In line with recent theoretical investigations~\cite{meleNonunitalNoise2024} we  expect that also other types of BP are  suppressed by the influence amplitude damping on real hardware.
For example, due to the contractive map at the heart of non-unital noise and the resulting noise-induced limit set, deep quantum circuits loose their expressive power.

However, the mere fact the a finite gradient is retained under amplitude damping, does not imply that such circuits are useful for solving a given task. 
As remarked above and in ref.~\cite{meleNonunitalNoise2024}, non-unital noise implies a finite and shallow effective circuit depth, which  limits the  expressive power, and therefore does not allow for the solution of complex problems.
In addition, only the variational parameters of the later layers  determine  the cost function values, and the influence of the  early layers is lost. 
This calls for  re-evaluating  insights related to parameter concentration and parameter transfer in variational quantum circuits~\cite{akshayParameterConQAOA2021,barreraTransferQAOA25,sakaiLinearQAOA2024,barreraLR-QAOA2025,pelofske2025evaluating}, where  warm-starting schemes based on transferring parameters or optimization-free schemes  might not lead to increased success probabilities. 
Yet, recent work suggest that non-unital noise and dissipative processes can be leveraged for useful computation~\cite{zapusek2025DissipationAlg} as well as in reservoir computing~\cite{nonunitalresource}.

It is still an interesting research topic how errors from non-unital noise channels like amplitude damping can be mitigated on near-term hardware, since popular schemes like dynamical decoupling, Pauli twirling and zero-noise extrapolation are not effective.
Error mitigation by symmetry verification~\cite{bonetSymmetry2018} is efficient as it can be done in postprocessing, but it is restricted  to specific problems. 
Approaches based on weak measurements~\cite{PhysRevLett.101.200401,PhysRevA.107.052409,yongWeakMeasure2014} also seem promising and might be easier to implement than full error correction schemes~\cite{fletcherAmpDamp2007,JayashankarAmpDamp2022,duttaAmpDamp2024}.
All of these are interesting research directions that we leave for future work.

\section{Methods}\label{sec:methods}
In this section we provide details necessary reproduce the numerical and experimetnal results in the mansucript.

\subsection{Circuit structure}\label{subsec:circuits_hardware}
\begin{figure}[h]
\centering
    \includegraphics[width=0.79\linewidth,trim=0 54mm 9.6cm 0mm,clip]{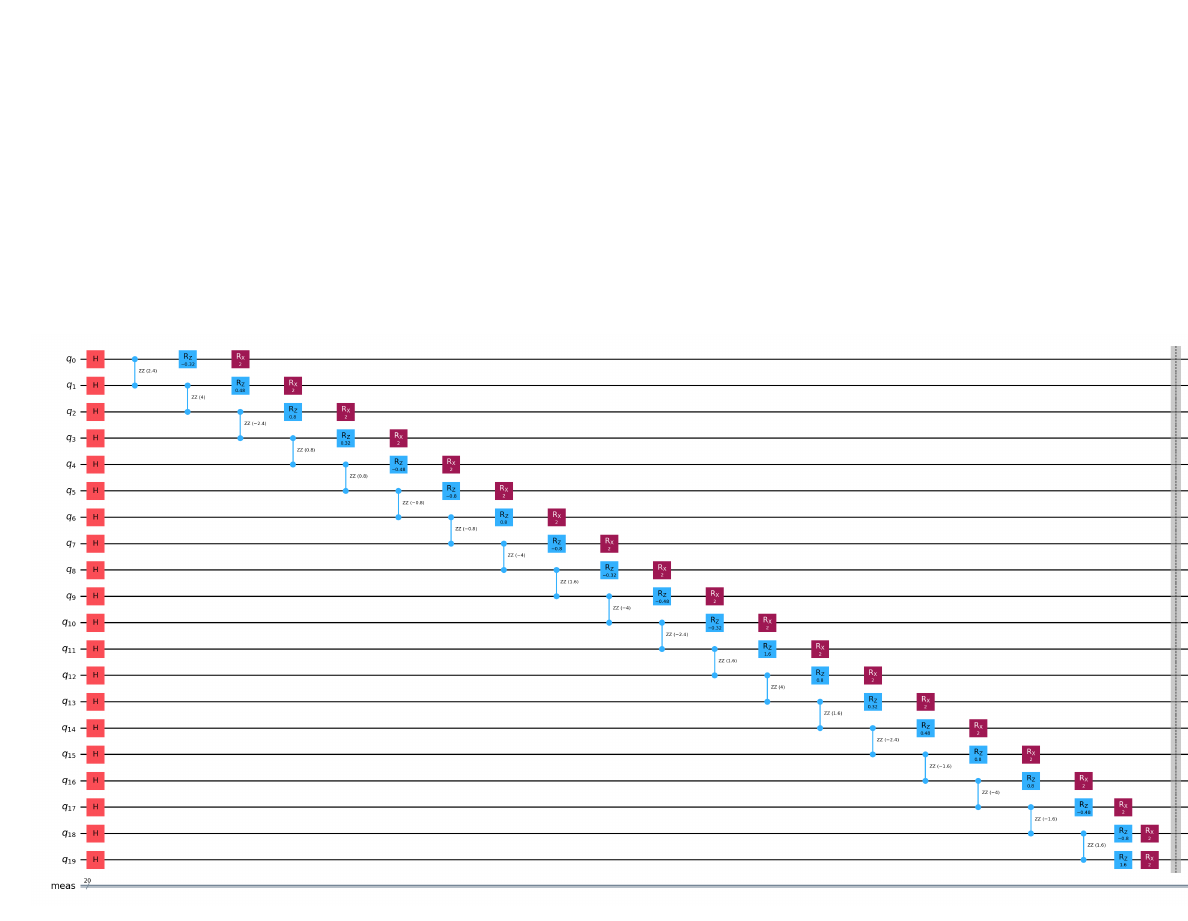} 
     \includegraphics[width=0.19\linewidth,trim=0 0mm 19.6cm 0mm,clip]{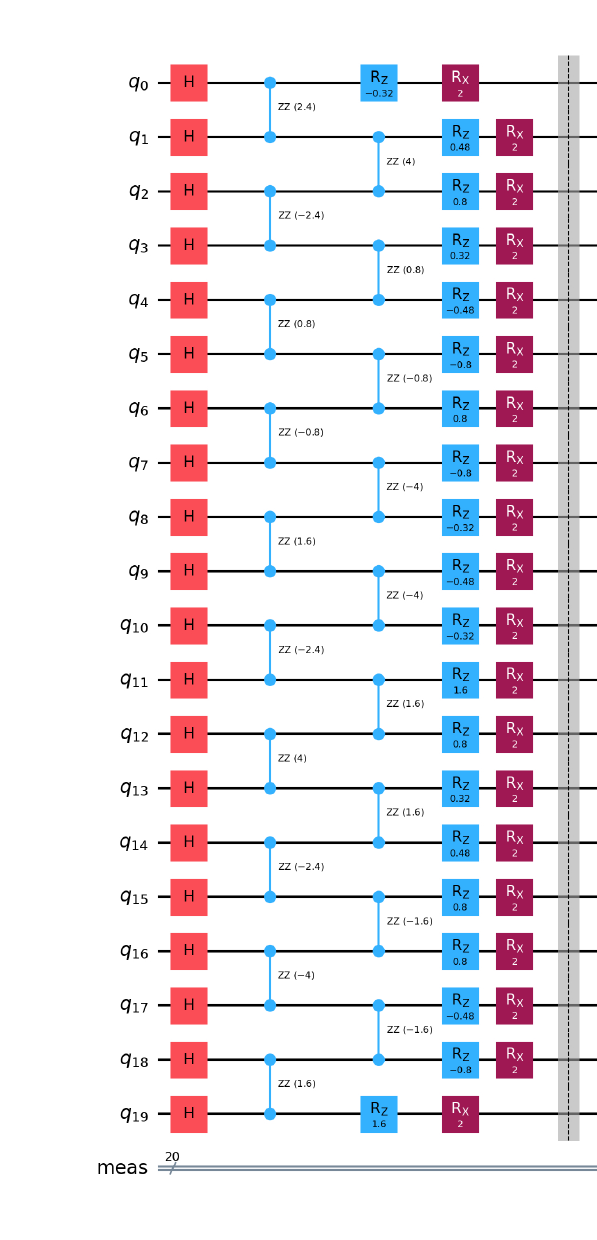}
\caption{Examples of the same logical $L=1$ single-layer circuit with $N=20$ qubits with two different transpilations.
Ladder structure (left) which is the default structure used in our experiment and parallel execution (right) which we termed 'short'.}
\label{fig:quantum-circuits-examples}
\end{figure}

\Cref{fig:quantum-circuits-examples} shows examples of the two types of quantum circuits that were ran on the IBM hardware, which only differ in their gate scheduling.
In the first implementation  (left panel)  the  two-qubit terms are executed in a ladder pattern which leads to a rather large circuit depth. 
This implementation, which  is generated by the naive formulation and transpilation of the circuits, is the one we used for almost all  of the experiments for this work.  

In the second implementation (right panel) commuting two-qubit terms are grouped and executed in parallel to minimize circuit depth. This variation of the circuit is termed 'short'. 

After transpilation to hardware, both circuit types have the exact same number of two-qubit gates, $2(N-1)L$,  where the factor of 2 comes from the fact that one $R_{zz}$ gate is realized  with two \texttt{ECR} gates on the Falcon platform. 
The number of single qubit gates varies slightly between circuits due to different possible optimizations of gate sequences for different parameter values, but is s always on the same order of magnitude for all circuits and scales approximately as $ 13.5\, NL $.

\subsection{Circuit depth and runtimes} 

\begin{figure}
    \centering
    \includegraphics[width=0.95\linewidth]{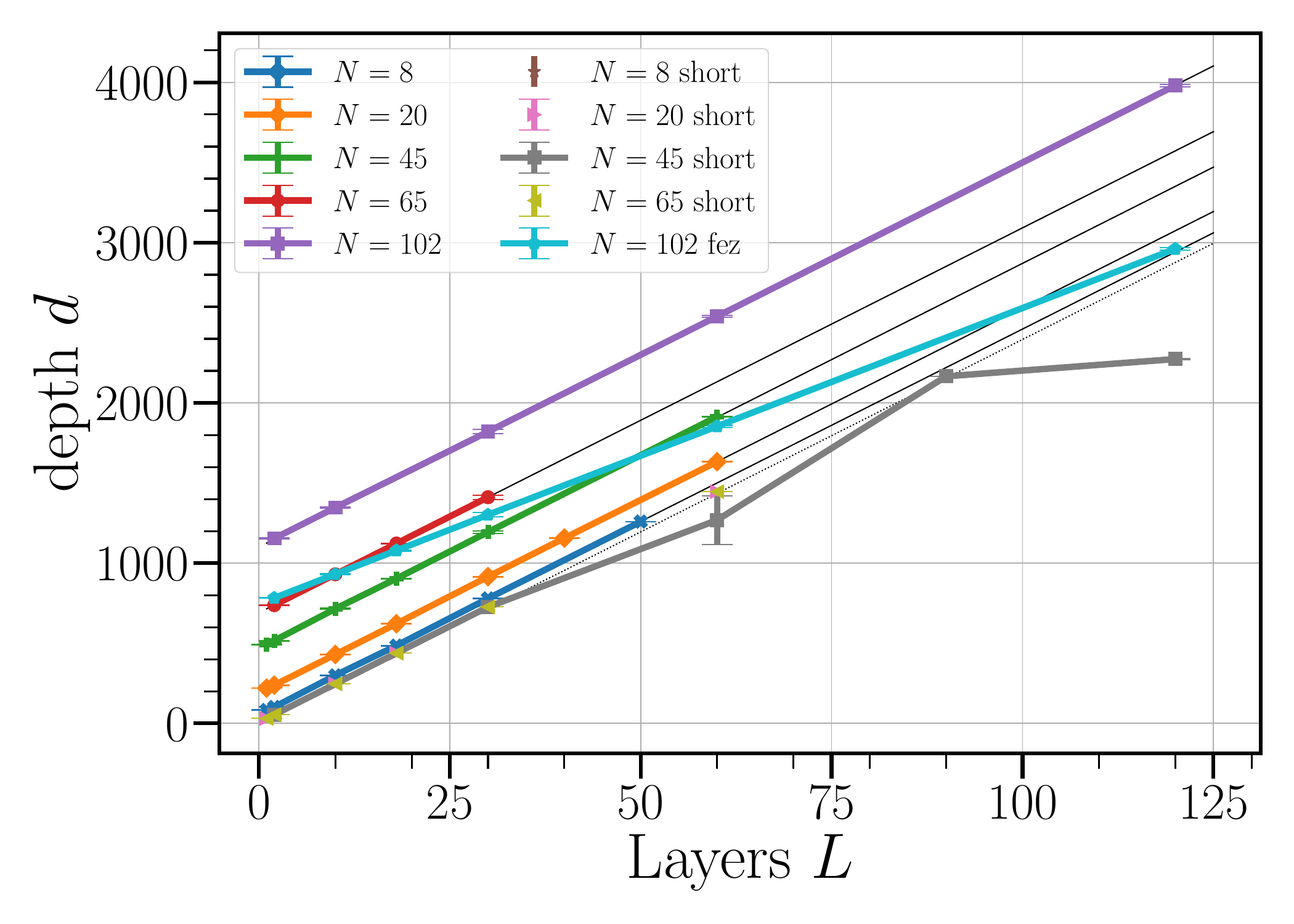}
    \\
    \includegraphics[width=0.95\linewidth]{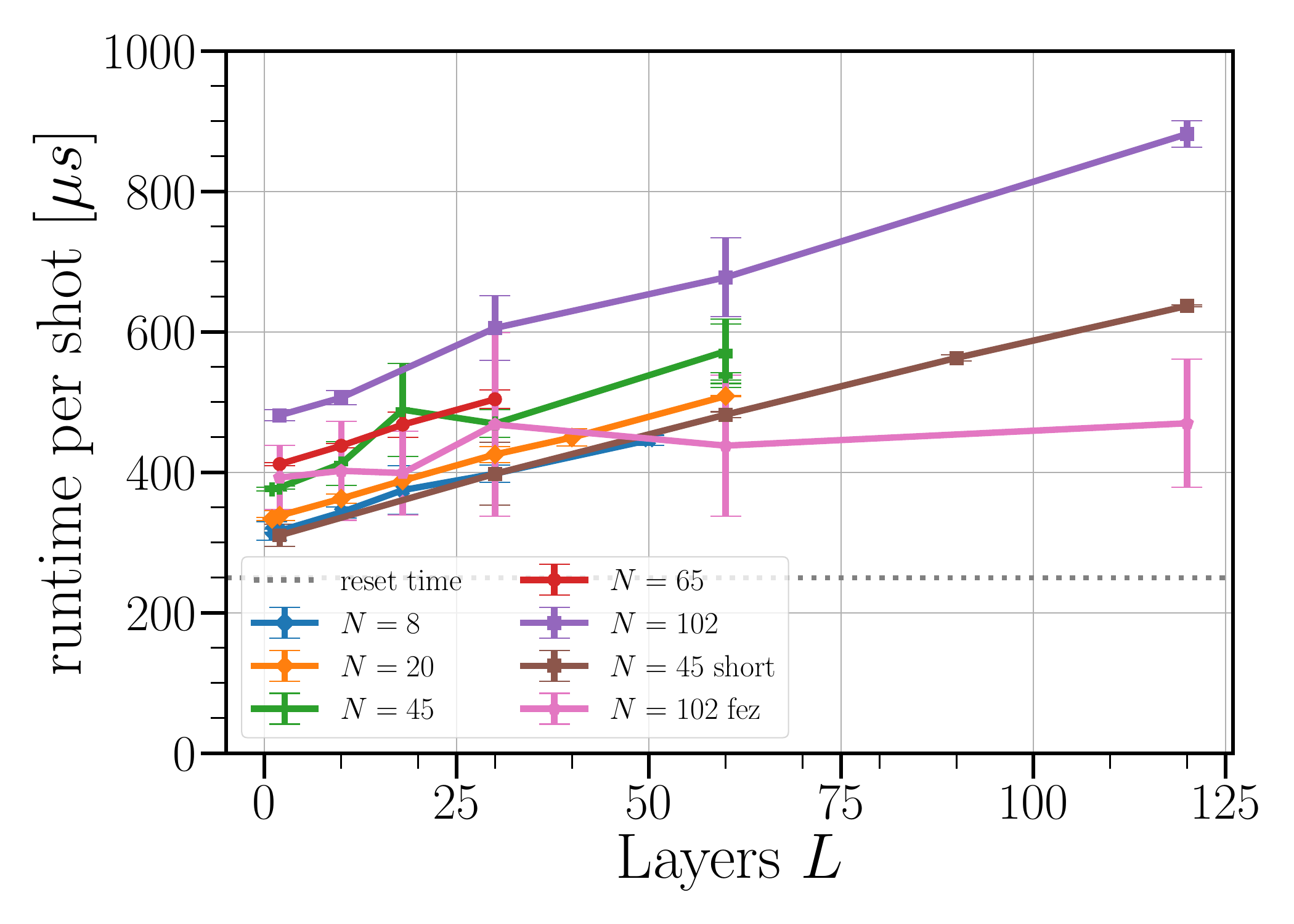}
    \caption{Circuit depth (top left) and circuit runtimes (other panels) as function of layers and depth for different numbers of qubits and from the \texttt{ibm\_brisbane} device unless otherwise stated in the legend.  } 
    \label{fig:depthStats}
\end{figure}

\Cref{fig:depthStats} shows the depth and runtime statistics of the circuits in the hardware experiments. 
The depth of the circuit (left panel) nicely follows the linear function    
\begin{align}
\label{eq:depth}
    d(N,L)&= a\, N + b\,  L + c
    \end{align}
 with $(a,b,c)=(11.1, 24.0, -29.6)$ for the ladder structure architecture and $(a,b,c)=(0, 24.0, -6.0)$ for the short circuits on the Falcon devices, and $(a,b,c)=(7.3, 18.5, 1.0)$ for the \texttt{ibm\_fez} device.

The measured runtimes of the circuits as function of the depth are shown in the right panel of ~\Cref{fig:depthStats}.
The runtime as function of the depth nicely follows the function 
\begin{align}
    \cirtimes^\texttt{Falcon}(d)=146ns\, d+305\mu s.
\end{align}
This runtime per depth of $146 ns$  coincides with estimating the runtime using the actual gate times from the hardware and  the statistical distribution of one and two-qubit gates per depth. 
In the ladder-like circuits we have in average about 2 two qubit gates and 10 single qubit gates per 12 depth, which gives an average runtime per depth of $t_\text{depth}\approx (10\cdot 60ns+2\cdot 660 ns)/12 = 160n s$. 

Accounting for the $250\mu s$ reset time between shots (\texttt{rep\_relay}), the net circuit runtime for one shot, i.e.\ the time for running the gates of the circuit and the measurement protocol, as function of number of qubits and circuit depth are:
\begin{align}
\label{eq:circ_runtime}
    \cirtimes^\texttt{Falcon}(N,L)&= 1.8\mu s\, N +3.3\mu s\,  L + 45\mu s\\
    \cirtimes^\texttt{Falcon,short}(N,L)&=\phantom{x.xx\mu s N + } 2.8\mu s  L + 61\mu s\\
    \cirtimes^\texttt{fez}(N,L)&= 0.24\mu s\,N + 0.61\mu s  L + 130\mu s 
\end{align}
The optimized circuit runtime is independent on $N$.
These function are used to determine  the circuit runtimes $\cirtimes$ in the main text.

\subsection{Calibration data }

\begin{figure}
    \texttt{ibm\_brisbane}\\
    \centering
    \includegraphics[width=0.43\linewidth]{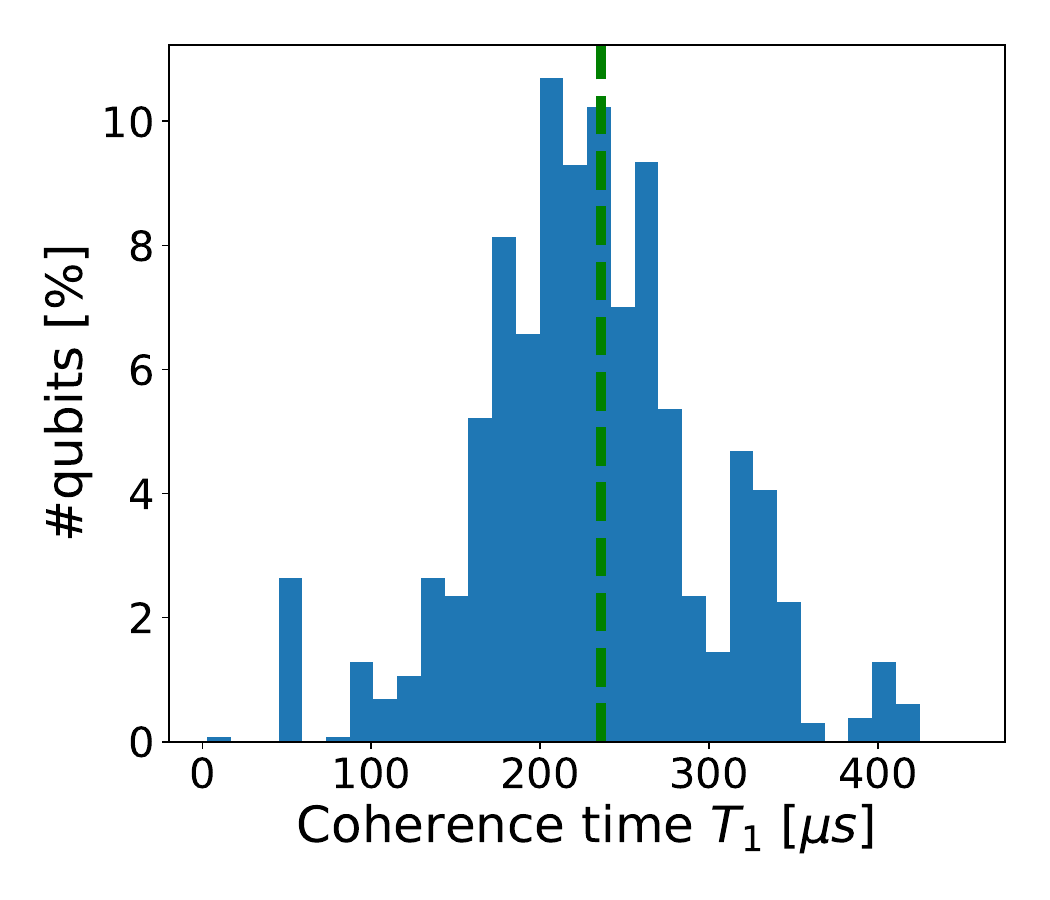}
    \includegraphics[width=0.43\linewidth]{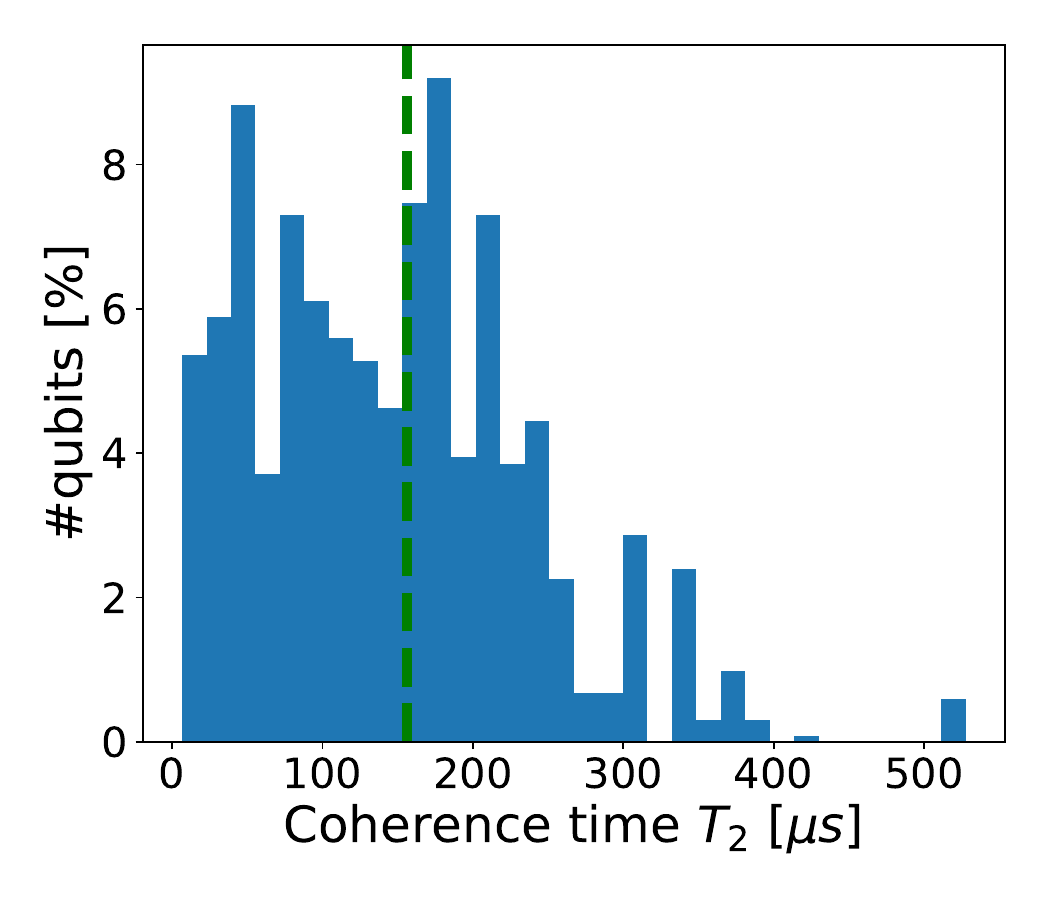}
    \\
    \includegraphics[width=0.43\linewidth]{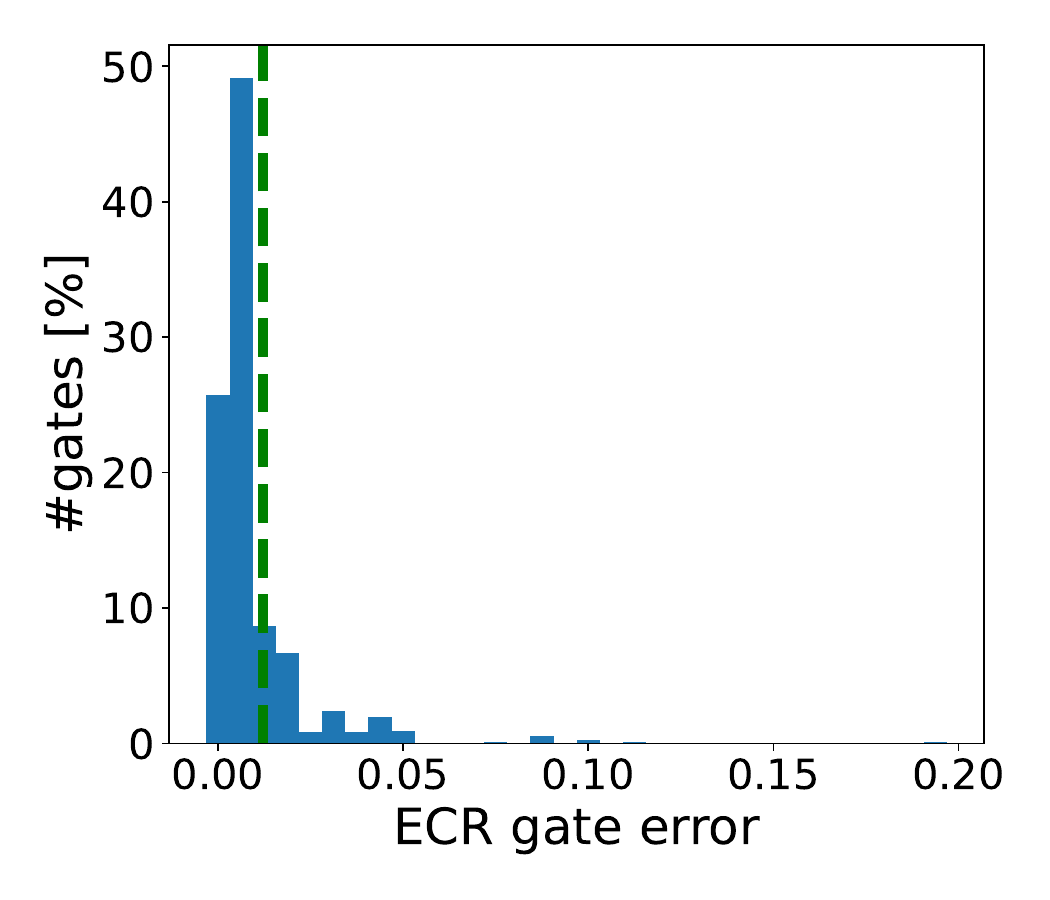}
    \includegraphics[width=0.43\linewidth]{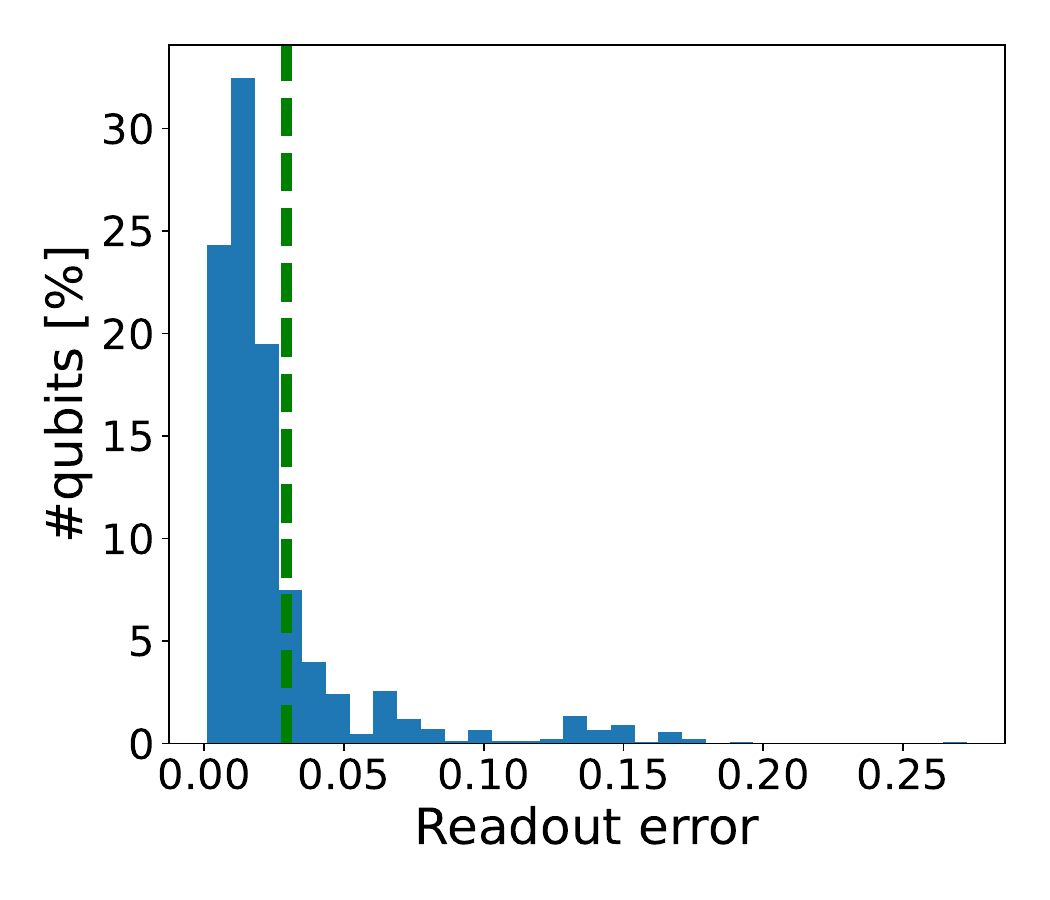}
    \\
    \texttt{ibm\_fez}\\
    \includegraphics[width=0.43\linewidth]{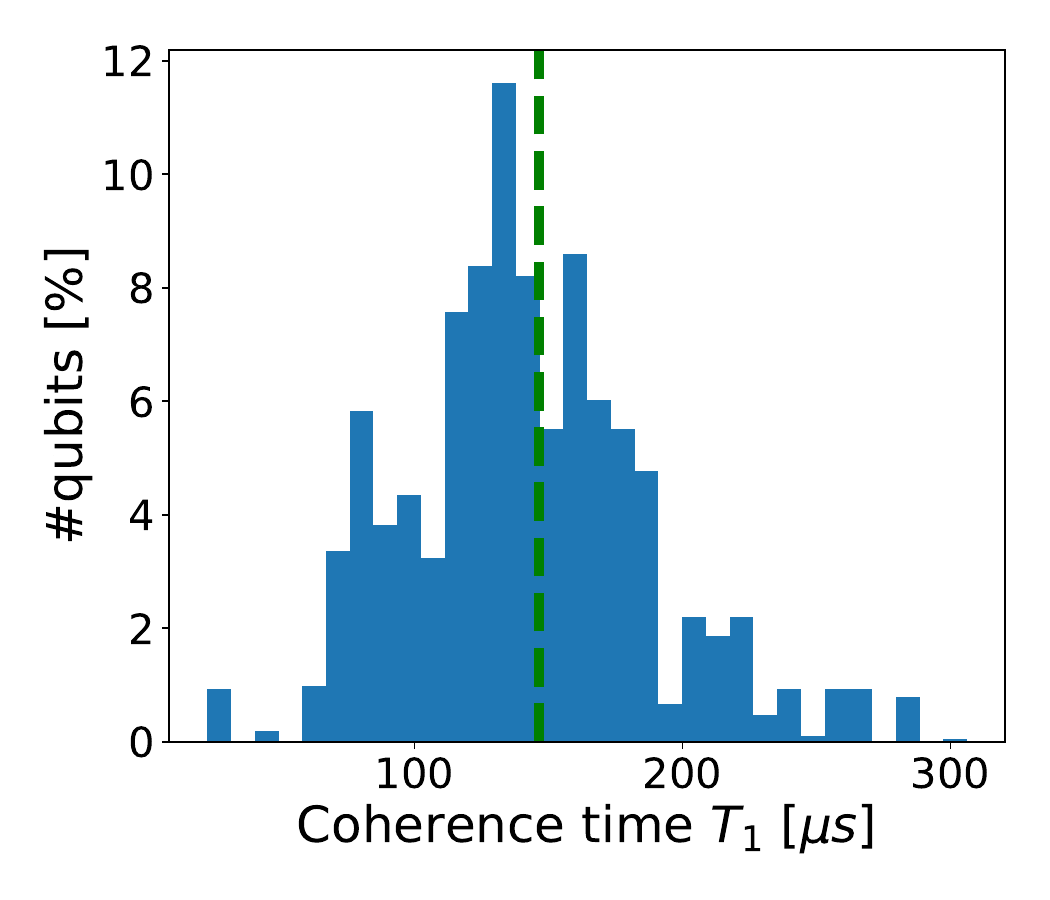}
    \includegraphics[width=0.43\linewidth]{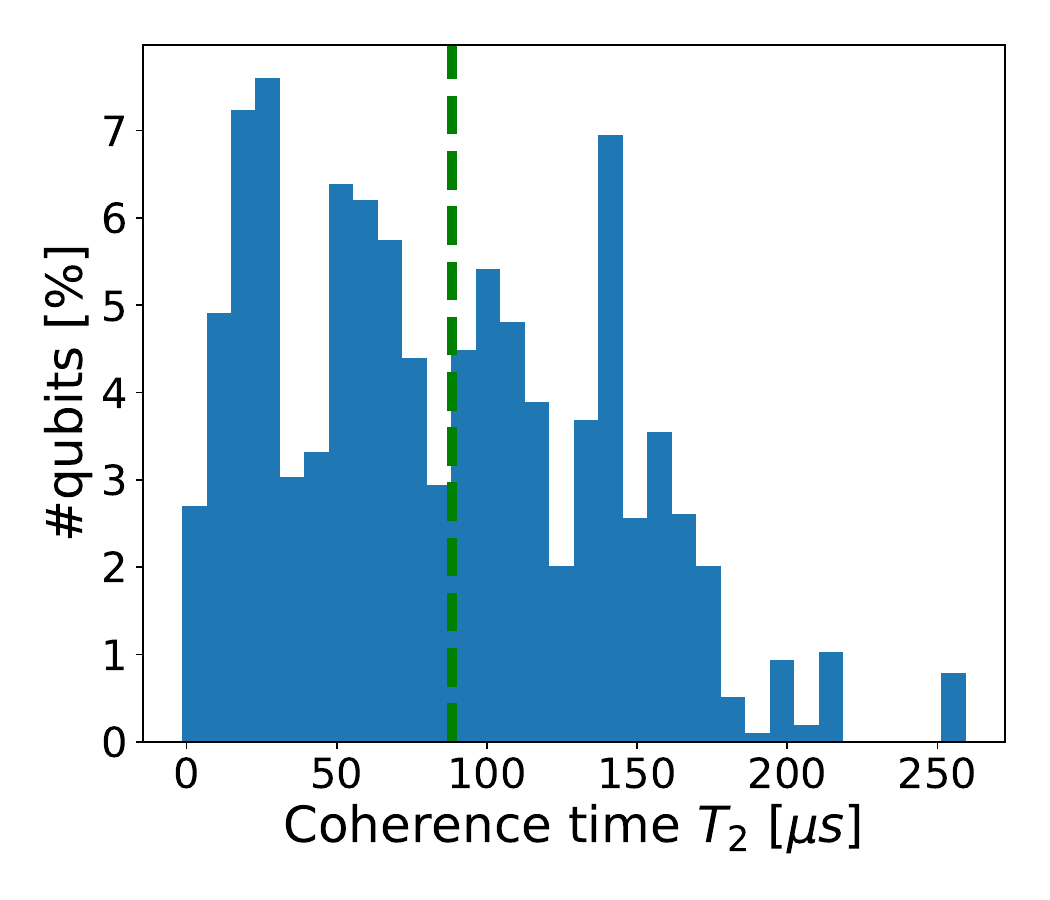}
    \\
    \includegraphics[width=0.43\linewidth]{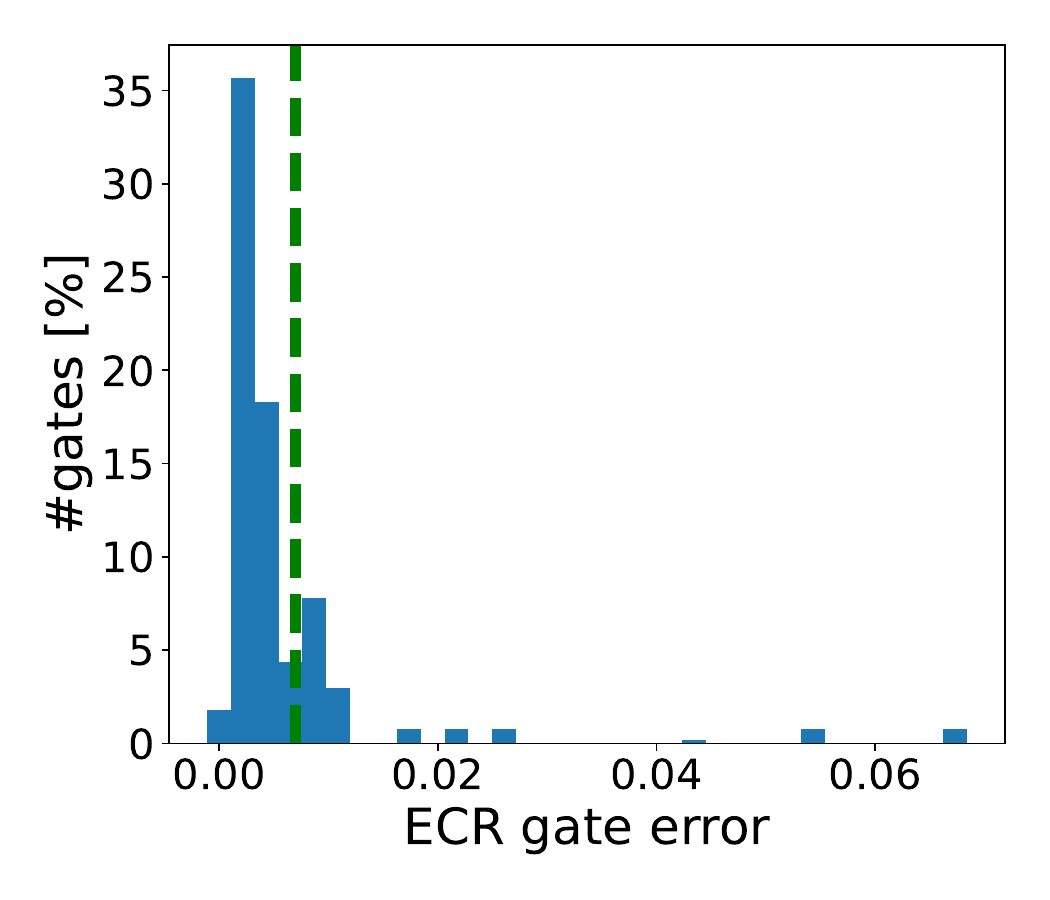}
    \includegraphics[width=0.43\linewidth]{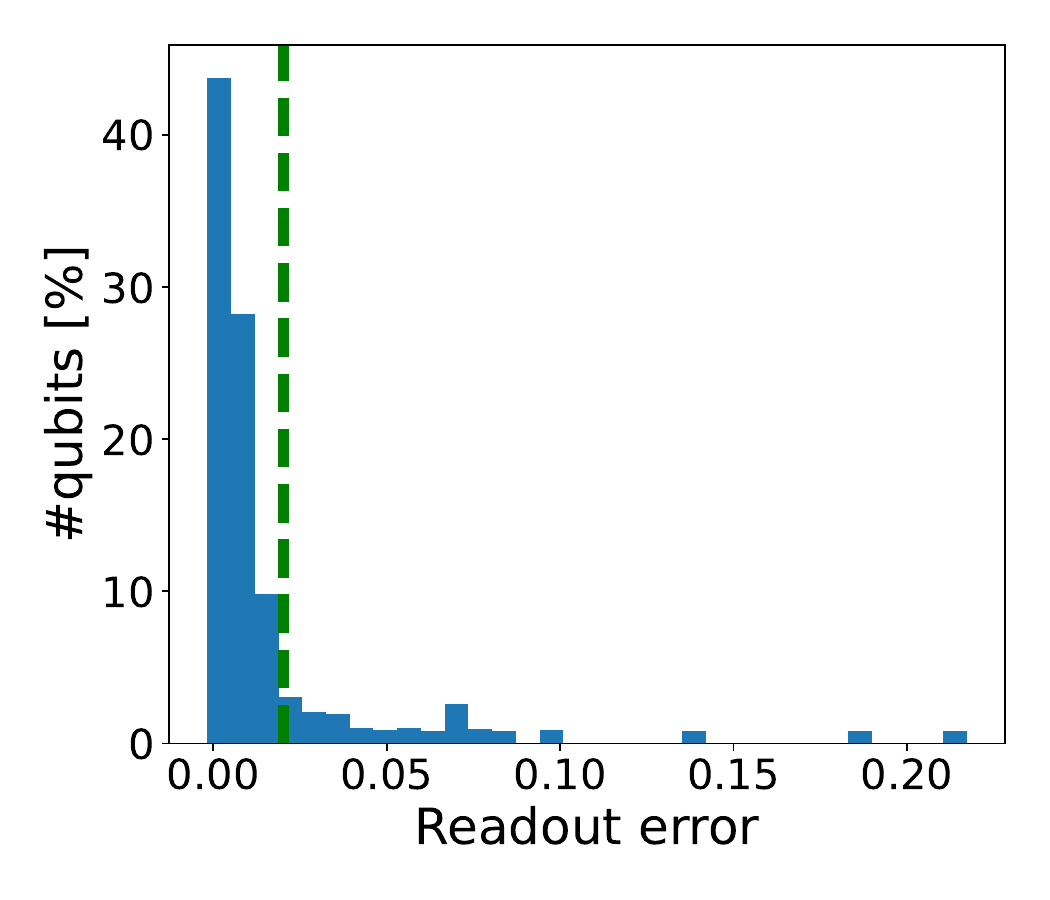}
    \caption{Statistics of the properties of the used hardware \texttt{ibm\_brisbane} (upper row) and \texttt{ibm\_fez} (lower row) for the $N=102$ qubits circuits. 
    The green vertical lines indicate mean values which are  are reported in Table~\ref{tab:calib_all}. 
    Note that all distributions very asymmetric and far from Gaussian distributions.  }
    \label{fig:calibration_brisbane}
\end{figure}

An overview on the distributions of the calibration data used for the experiments on \texttt{ibm\_brisbane} and \texttt{ibm\_fez} for the $N=102$ qubit circuits are shown in Fig.~\ref{fig:calibration_brisbane}. 
The  estimated mean values are shown in \Cref{tab:calib_all}  for all different platforms.

\begin{table*}
    \centering
\begin{tabular}{|l|c|c|c|c|}
   \hline 
 device  & $\langle T_1\rangle$ [$\mu s$] & $\langle T_2\rangle $  [$\mu s$] & $\langle \texttt{readout\_error}\rangle $ [\%]& $\langle \texttt{ERC\_error} \rangle$  [\%]\\ \hline  
\texttt{ibm\_brisbane} &$236\pm67 $& $156\pm 93$ & $3.0\pm 3.0$ & $1.2\pm1.4$  \\\hline
\texttt{ibm\_kyiv}&$302\pm 91$& $155\pm 113$ & $2.2\pm 2.7$ & $1.4\pm1.4$  \\\hline
\texttt{ibm\_sherbrooke} &$310\pm90 $& $181\pm 136$ & $3.3\pm 3.4$ & $1.2\pm1.4$  \\\hline
\texttt{ibm\_fez} &$146\pm46 $& $88\pm 54$ & $2.0\pm 3.2$ & $0.7\pm0.9$  \\\hline
\end{tabular}
    \caption{Summary of the calibration data statistics from all runs on a specific device. The errors are the standard deviation calculated from the data. 
    Note that this is not a proper measure for error and only indicates the order of magnitude for the variations since  the  distributions are very  non-Gaussian. 
    }
    \label{tab:calib_all}
\end{table*}

\subsection{Robustness of ICLA}\label{sec:robustICLA}

\begin{figure}
    \includegraphics[width=0.95\linewidth]{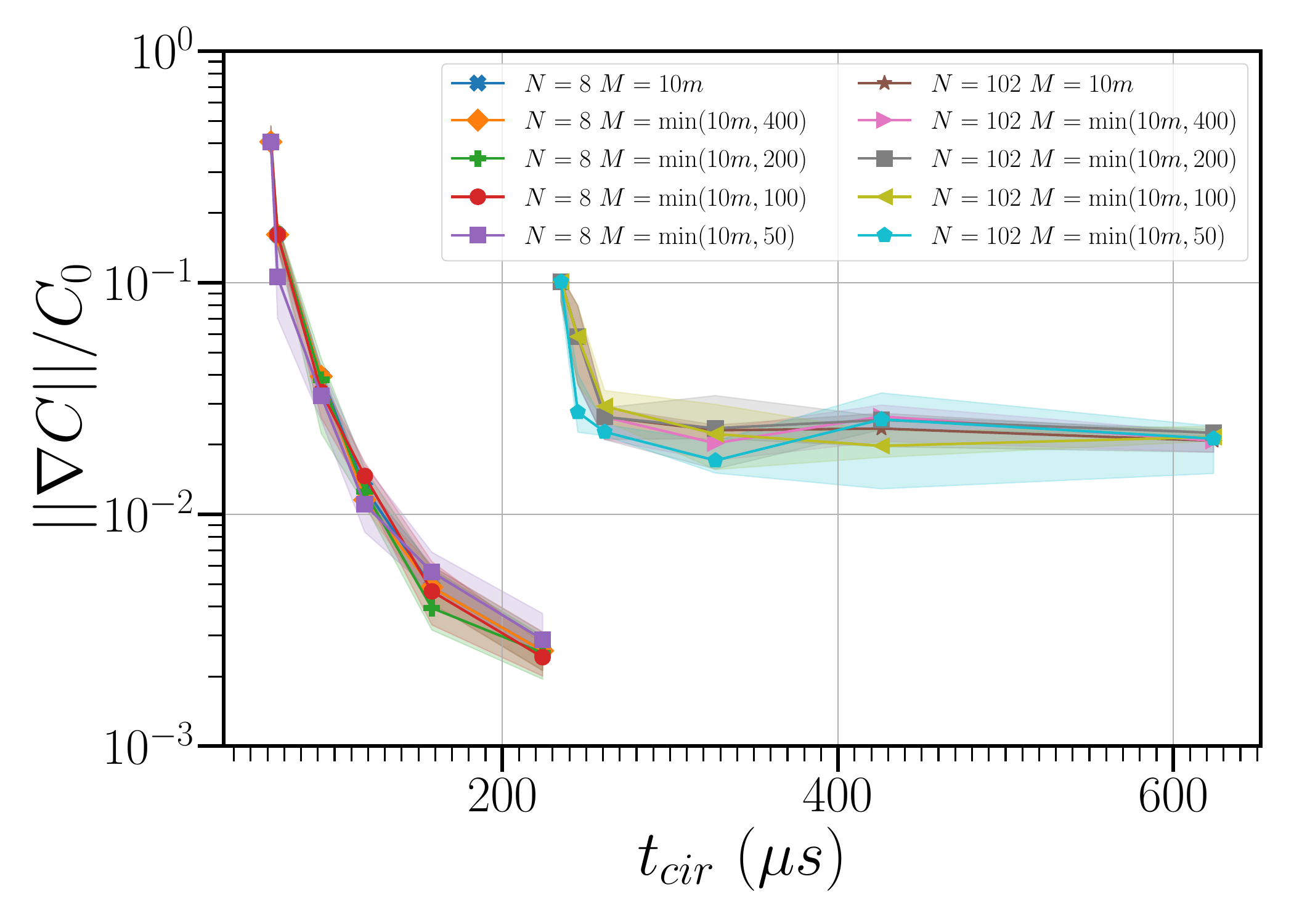}
    \caption{ Dependence of the ICLA results on the number of sample points $M$ used for the cost function landscape for $N=8$ and $N=102$ qubits circuits ran on the \texttt{ibm\_brisbane} device.  }
    \label{fig:ical_robust}
\end{figure}

In order to reduce the required number of hardware experiments we limit the cost function landscape to $M=\min(10m,200)$ sample points and calculate the ICLA from this reduced set.
In order to justify this reduction  compared the original ICLA~\cite{perez-salinas2024analyzing}, 
we calculated the ICLA  with various reductions. 
 \Cref{fig:ical_robust} shows the results for circuit with $N=8$ and $N=102$ and for various 
 maximal sizes for the cost function landscape.

 As can be observed the ICLA is very robust and all results are within their respective error bars. 
 Only for a maximal size of $M=50$ and for $N=102$ qubits larger variations in the result occur. 

This empirically observed robustness of the ICLA  constitutes a major reduction in the required  computational resources. 
For deep circuits, this implied a substantial  reduction in the number of required cost function evaluations, for example,  from $M=20L=1200$ to $200$ for $L=60$.

\subsection{Estimating cost function values}\label{subsec:costFunctions}
This section provides  details on how the cost functions values are obtained from the experiments and analyze the influence of shot noise and different calibrations. 

\begin{figure}
    \centering
    \includegraphics[width=0.96\linewidth]{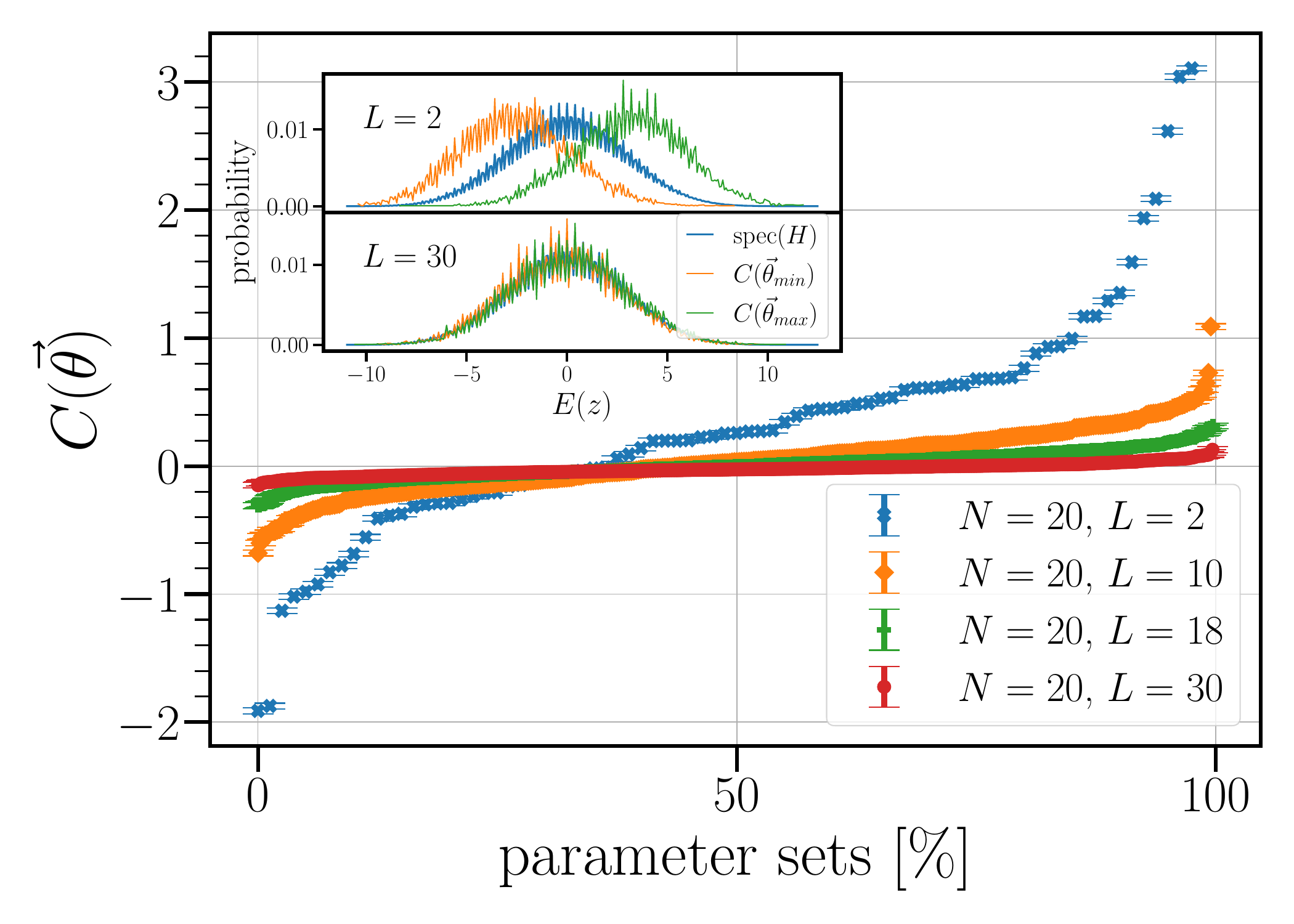}
    \\
    \includegraphics[width=0.96\linewidth]{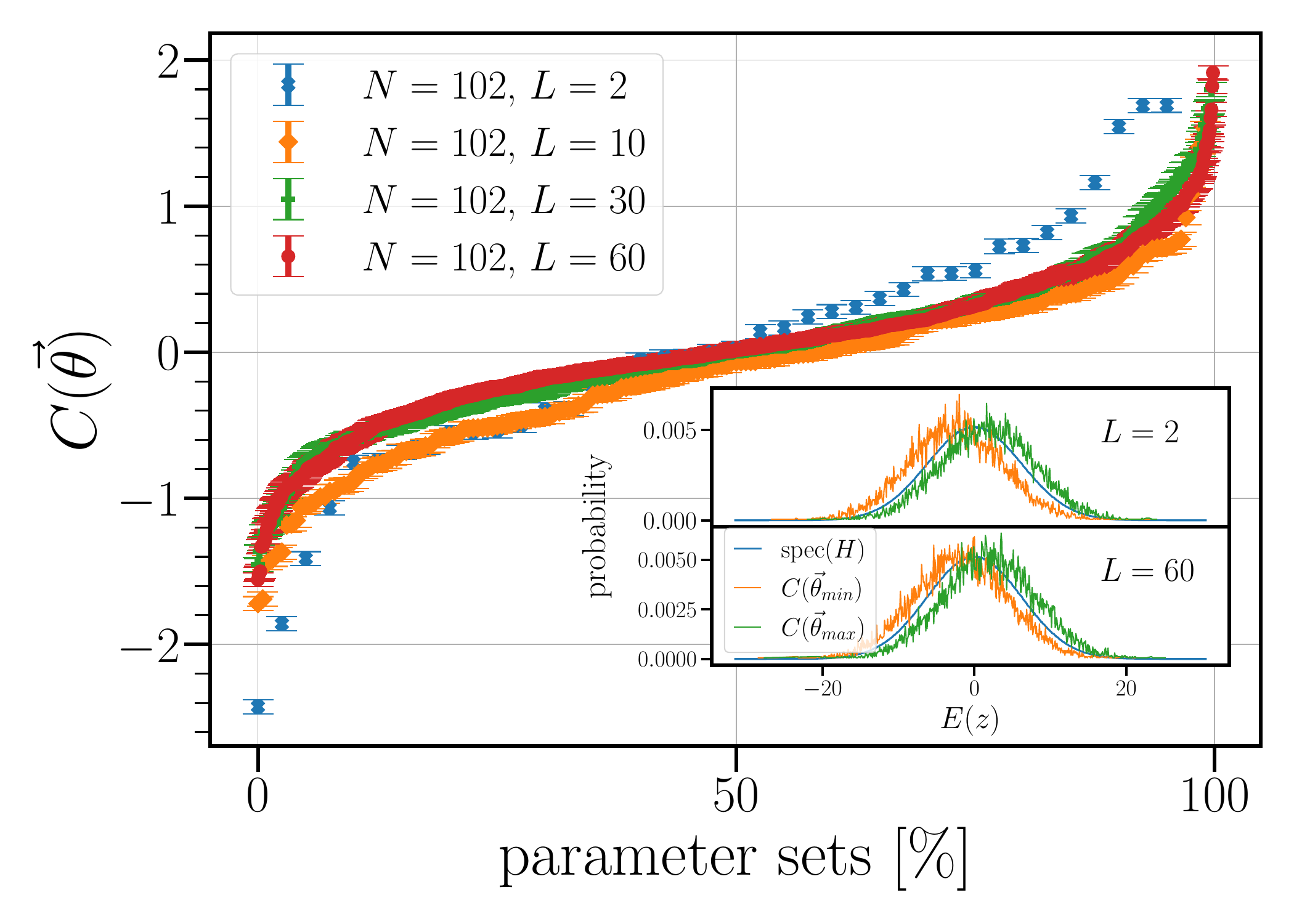}
    \caption{Sorted cost values for $N=20$ (upper) and $N=102$ (lower) from the experiments of the \texttt{ibm\_brisbane} device. 
    The error bars indicate the variation expected for to shot-noise with $R=16384$ shots used here. 
    The insets show cost  spectra for parameter values producing the minimal (orange) and maximal (green) cost function values for small (upper) and large (lower) number of layers. The blue curve is the spectrum of the Hamiltonian.
    }
\label{fig:EnergySpectra}
\end{figure}

\Cref{fig:EnergySpectra} shows the mean cost values as obtained with Eq.~\eqref{eq:samp-cost} by an average over $R=16384$ circuit samples  for small ($N=20$) and large ($N=102$) circuits, and for several different circuits depths as function of the variational parameters  used for the ICLA.
The plots reveal a clear structure of the cost values showing a characteristic S-shape. 
Most cost values are  small and close to zero, while there are some circuit parameters for which produce large (positive and negative) cost  values reaching up to the scale of the Hamiltonian energies. 
For small  number of qubits ($N=20$, left panel) the absolute values of the cost values diminish  with increasing circuit depth $L$ and seem to approach the same value regardless of the circuit parameters. This leads to the vanishing of  the gradient information as observed. On the other hand, for large number of qubits ($N=102$, right panel) the cost function values slightly decrease from $L=2$ to $L=10$ layers, but after that  retain the same S-shape. This corresponds to the gradient signal flattening out at a nonzero value as observed.  

We calculate the cost function values by taking  $R$ measurement shots of the quantum circuit with the corresponding circuit  parameters $\vtheta$ and take the mean over the cost of each bitstring (Eq.~\eqref{eq:costSingle}). Therefore, it is illustrative to investigate the distribution of cost function values leading to  the mean values.
These are also shown in~\Cref{fig:EnergySpectra} as insets for several configurations. 
First, it can be observed that the width of the distributions are rather large and determined by the spread of the  spectrum of the Hamiltonian, i.e.\ $C_0=\sqrt{\langle {{\rm H}_C}^2\rangle}
=\frac{\sqrt{N}}{2}\sqrt{\tfrac{N-1}{N}\overline{J^2} +\overline{h^2} }\sim\sqrt{N}
$, where  $\overline{J^2}$ and $\overline{h^2}$ are the mean of the squared parameters of the Hamiltonian Eq.~\eqref{eq:ising}.  
The spectra of the Hamiltonians are also shown in the figure as the blue curves. 
However, depending on the circuit (width and depth), the spectra for specific circuit parameters show a pronounced shift, which are of course related to the mean cost function values shown in the main panels.

These plot reveal, that while sampling from a quantum circuit produces solutions with from a  wide range of energies, the circuit parameters have a profound influence by shifting the center of the  distribution.

\subsection{Noise influences}\label{subsec:noise_influence}
\begin{figure}
    \centering
    \includegraphics[width=0.95\linewidth]{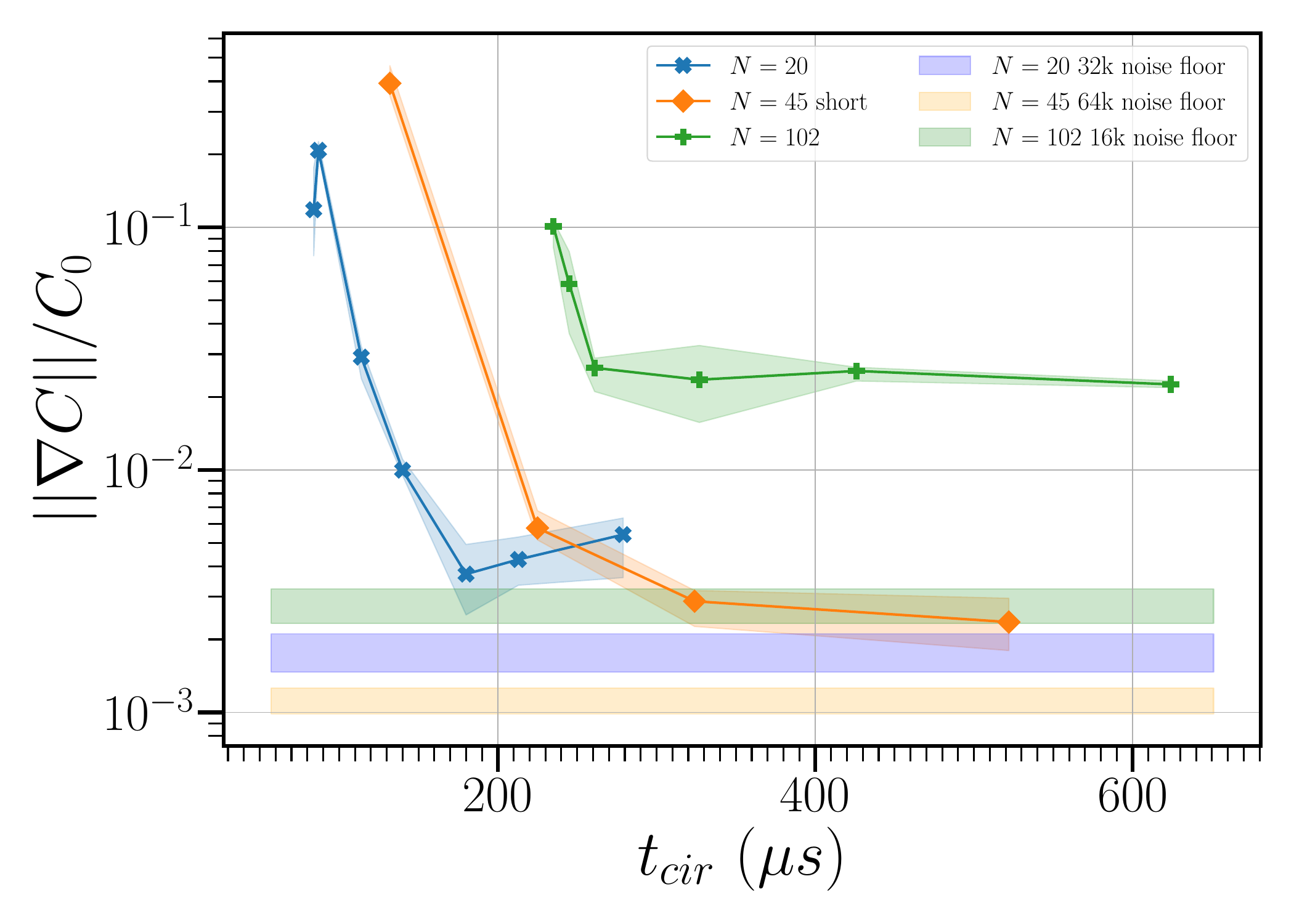}
    \caption{Gradient signal for different configurations and the corresponding shot noise limits. 
    We calculated the  
    }
    \label{fig:shotNoise}
\end{figure}

The finite gradient signals as observed in our work could also be the result of pure random noise with the appropriate amplitude (instead of being a determined by variational circuit parameters, as we argue).
The first candidate for such noise would be shot noise due to finite number of sampling shots. 
The shot-noise variations are indicated  by the error bars in \Cref{fig:EnergySpectra} and are rather small due to the large number of shots used. 
To make this more explicit, we also calculated the ICLA shot noise floor by generating cost function landscapes where the actual cost is replace by random sampling of the spectrum of the Hamiltonian, and then calculating the ICLA from this cost function landscape.
The results are shown in~\Cref{fig:shotNoise} for three different examples.
It can be clearly seen that the noise floor is substantially below the actual sizes of the gradient signals.  

\begin{figure}
    \centering
    \includegraphics[width=0.98\linewidth]{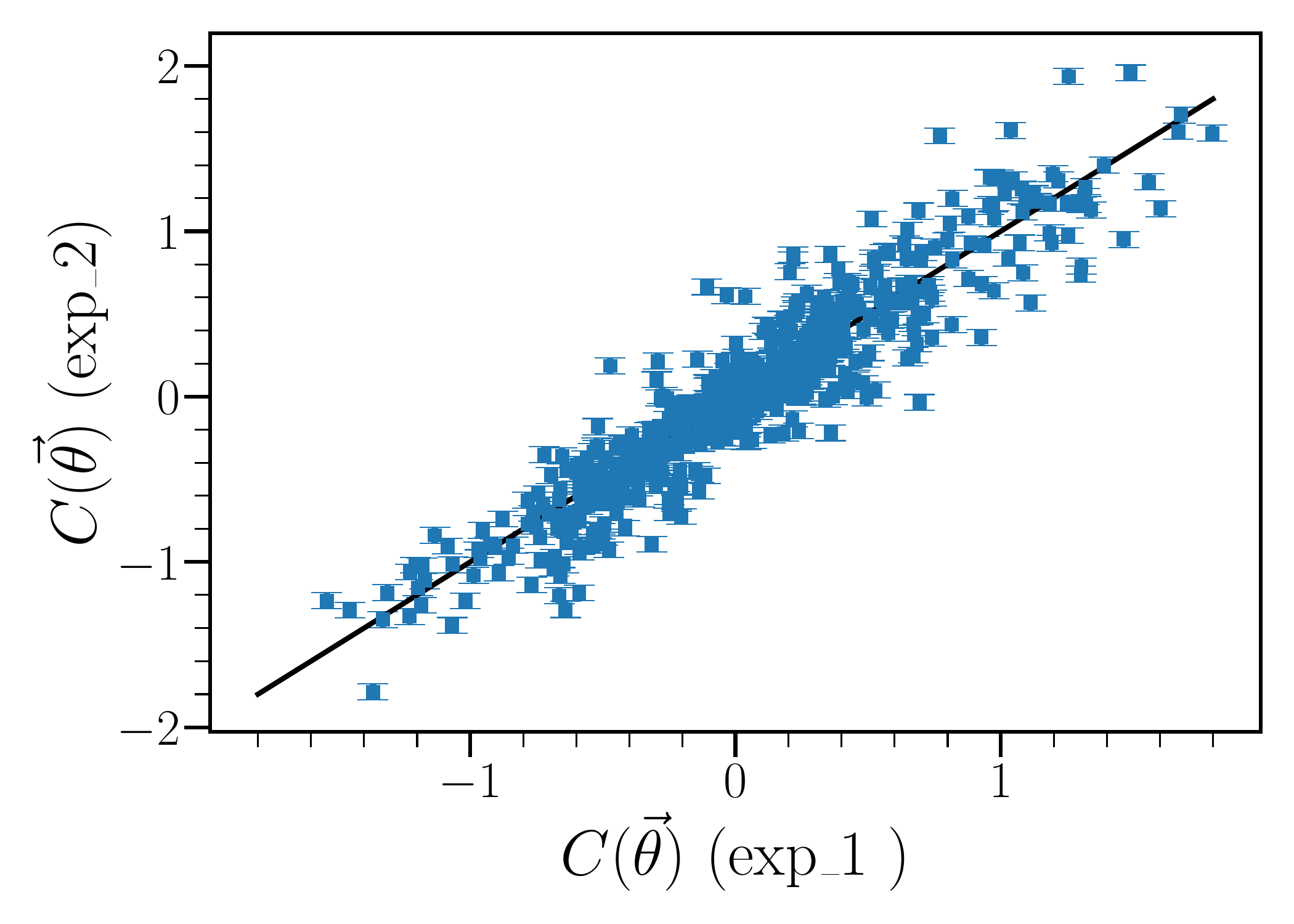}
    \caption{Mean cost function values for the exact same circuits for $N=102$ and $L=30$ with parameters evaluated on the \texttt{ibm\_brisbane} twice, but with four days between evaluations between experiment 1 and 2.
    }
    \label{fig:CostCorrHardware}
\end{figure}

Even though we targeted to evaluate one specific setting circuit type and size ($N$ and $L$), for several experiments  there were different calibrations used for different parameter sets.  
In order to evaluate the influence of different calibrations, we estimate mean cost values from two different experiments.
We ran circuits with $N=102$ qubits and $L=30$ layers on \texttt{ibm\_brisbane} for 600 different circuit parameters, $\{\vtheta_1,\dots,\vtheta_{600}\}$ and estimated the mean cost function values for each parameter set, $C^\texttt{exp\_1}_M(\vtheta_p)$ ($p=1,\dots,600$) with $M=16384$ shots.
Then, we performed the second experiment four days later where we ran the exact  same circuits on the same machine with the same $600$ parameters and estimated again the same mean cost function values, $C^\texttt{exp\_2}_M(\vtheta_p)$ ($p=1,\dots,600$, $M=16384$).     
We made sure that always the exact same physical qubits were used for all circuits, as well as the exact mapping from logical to physical qubits. 
However, the calibrations were slightly differing between runs, see Table~\ref{tab:calib}. 
\begin{table*}
    \centering
\begin{tabular}{|l|c|c|c|c|c|}
   \hline 
   & $\langle T_1\rangle$ & $\langle T_2\rangle $  & $\langle \texttt{readout\_error}\rangle $ & $\langle \texttt{ERC\_error} \rangle$  & Num.\ circs\\ \hline  
exp\_1, calib\_1 &$237 \mu s$& $157\mu s$ & 2.88\% & 3.18\% & 502  \\ 
exp\_1, calib\_2 &$237 \mu s$& $157\mu s$ & 3.02\% & 3.26\% & 81 \\ 
exp\_1, calib\_3 &$237 \mu s$& $157\mu s$ & 3.02\% & 3.18\% & 17  \\\hline
exp\_2, calib\_1 &$236 \mu s$& $157\mu s$ & 3.08\% & 3.06\% & 600  \\ 
\hline
\end{tabular}
\label{tab:calib}
   \caption{Statistics of the hardware calibration used for the two different experiments shown in Fig.~\ref{fig:CostCorrHardware}. 
   All values are calculated from the actual physical qubits used in the experiments which were the exact same for all circuits.}
\end{table*}

In Figure~\ref{fig:CostCorrHardware} we show the estimated mean cost values from both experiments, $C^\texttt{exp\_1}_M(\vtheta_p)$ and $C^\texttt{exp\_2}_M(\vtheta_p)$.  
The plot reveals that the cost function estimations of the two runs are highly correlated, and reproducible.
The observed variation between the two experiments is larger than the variation expected from the shot noise (error bars). 
But due to the large correlations we conclude that the signal we observe are not produced by random variations of the cost functions, but instead are the results of the deterministic influence of the variational parameters.
However, as we already discussed in the main text, different calibrations have an influence on the results.

\subsection{Density matrix analysis}\label{subsec:density-matrices}
To complement the experimental analysis on IBM quantum devices, we performed numerical simulations to isolate the role of specific noise channels in the emergence of noise-induced barren plateaus.

We simulated the same variational circuits used in the hardware experiments, after transpilation to the native gate set of the target IBM devices. Due to the exponential cost of density-matrix simulations, we restricted the system size to $N=8$ qubits.

Simulations were performed using qiskit considering two noise models: (i) amplitude damping noise, parameterized to emulate $T_1$ relaxation, and (ii) depolarizing noise, applied independently to each gate and qubit. We analyze the eigenvalue distributions of the final density matrices as a diagnostic of the underlying optimization landscape.

Depolarizing noise is expected to drive the output state toward the maximally mixed state $\rho =\frac1{2^N}\mathbb{I}$,
characterized by a eigenvalue distribution which has only one peak at $\rho_n=1/2^N$ with multiplicity $2^N$. 
Such spectral concentration is consistent with the emergence of a noise-induced barren plateau, where gradients vanish due to loss of state distinguishability.

Amplitude damping noise, by contrast, is non-unital and introduces a projective map in state space which does not allow for anti-concentration in the  circuit output map \cite{feffermanNonUnitalNoisePRXQ2024}. 
As a result, the dynamics converge toward a noise-induced limit set \cite{Pat2025beyondunitalnoise} rather than the maximally mixed state, leading to a saturation of the average gradient at a finite value, in agreement with our experimental observations.

\begin{figure}[t]
    \centering
    \includegraphics[width=1.0\columnwidth]{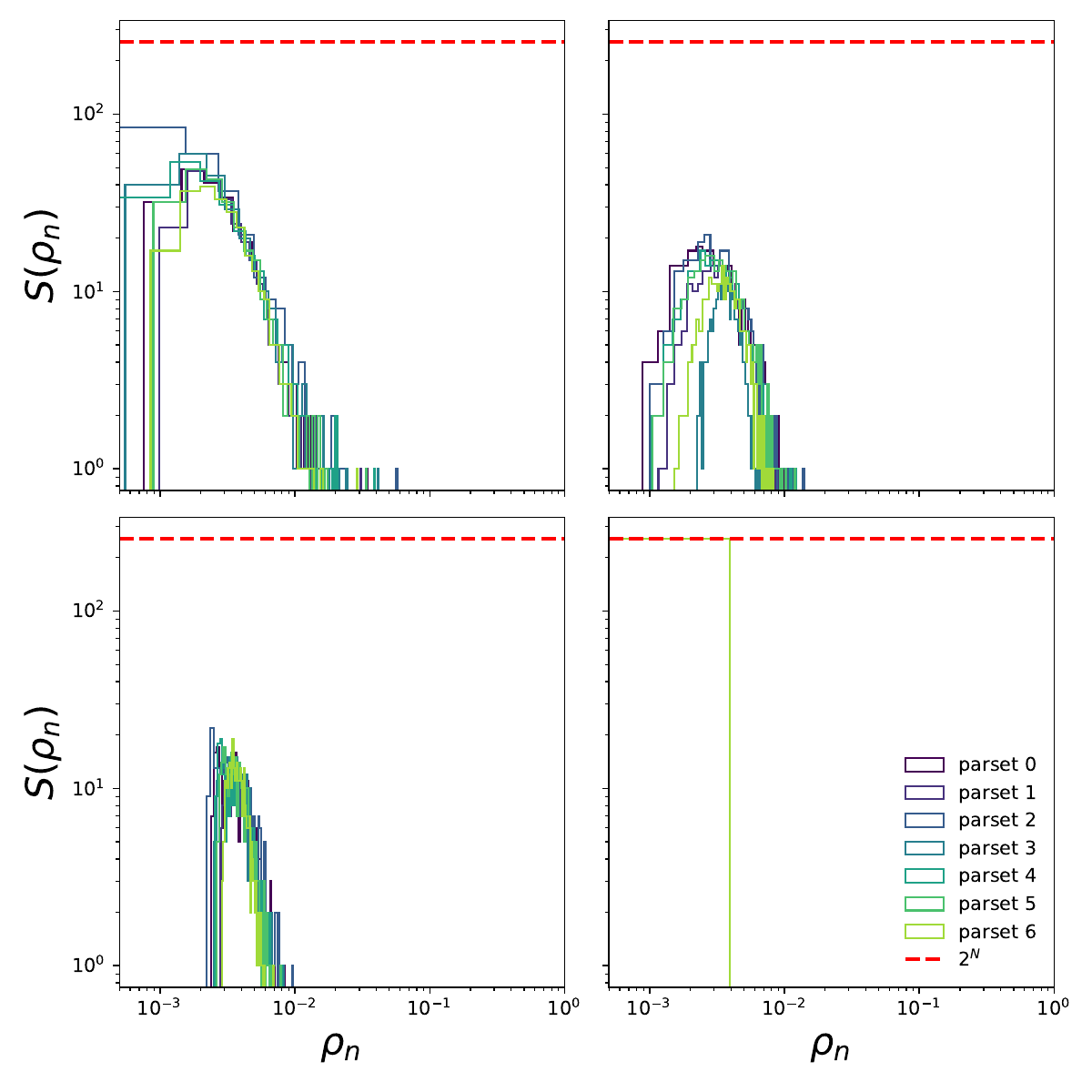}
    \label{fig:eigenvaluedistribution}
    \caption{
    Eigenvalue distribution of the simulated density matrices for $N=8$ qubits for $L = 10$ (left column) and $L= 50$ (right column) layers.
    (top-row) Under amplitude damping noise (error probability $p_1 = 0.05$), the persistence of dominant eigenvalues indicates convergence toward a noise-induced limit set. 
    (bottom-row) Under depolarizing noise (error probability $p_2 = 0.05$ for two qubit gates), the spectrum concentrates around $1/2^N$, consistent with convergence toward the maximally mixed state.}
    \label{fig:combined_spectra}
\end{figure}

Figure~\ref{fig:combined_spectra} displays the eigenvalue distributions of the density matrices for two different noise model and circuit depth.
Top row show the distribution under amplitude damping noise. 
The spectrum remains non-uniform, with a small number of dominant eigenvalues persisting even at finite noise strength.
Conversely, the bottom row shows shows the results under depolarizing noise, where the spectrum concentrates around the uniform value $1/2^N$, indicating convergence toward a maximally mixed state.

Crucially, by observing the distribution of the eigenvalues under amplitude damping and realizing that it remains structurally similar across different sets of variational parameters, we can understand the physical reason for the appearance of noise-induced limit sets. 
This spectral robustness suggests that non-unital noise contracts the state space into a specific sub-manifold (the noise-induced limit set) that is determined by the  noise channel. 
However, the limit set is still influenced by the variational parameters, which leads to finite cost function variations.   

The rough characteristics of the spectrum in the case of amplitude damping can be understood by a very simple model.
We assume we start in a pure state of $N$ spins which are all in the excited state, $\rho_0=\ketbra{1}{1}^N$.
Under amplitude damping with probability $p$ for each qubit, this state decays into 
\begin{align}
    \rho_0=\ketbra{1}{1}^N   \Rightarrow \sum_{n=0}^{N}p^n(1-p)^{N-n}\sum_{\substack{z\in\{0,1\}^N\\ \braket{z}{0^N}=n}}\ketbra{z}{z}
\end{align}
where the second sums runs over all states with where exaclty $n$ qubits are in state  $\ket0$, i.e. have exactly $n$ $0$s in the bitstring $z$.
The eigenvalues and corresponding multiplicities for this density matrix are
\begin{align}
    \rho_n& =p^n(1-p)^{N-n}\\
    S(\rho_n)&=\binom{N}{n}=\frac{N!}{(N-n)!n!}.
\end{align}
This is very similar to overall the form observed for the amplitude damping of~\Cref{fig:combined_spectra} (upper right panel). 

Although limited to a small system size, these simulations highlight how unital (depolarizing) noise drives the system towards the maximally mixed state and induces barren plateaus, while non unital (amplitude damping) noise preserves spectral structure and leads to a noise-induced limit set, which leads to finite average gradients.


%\bibliography{references}

\section*{Acknowledgments}
The authors would like to thank Phillip Hauke, Adri{\'a}n P{\'e}rez-Salinas, Stefano Polla, Daniel Lidar, and Phattharaporn Singkanipa for stimulating and fruitful discussions.
The authors also would like to thank Bernhard Sendhoff, Jens Schmuedderich, Heiko Wersing, Avetik Harutyunyan, and the Honda Research Institute network for support in this project.
A.B.\ acknowledges funding from the Honda Research Institute Europe GmbH, Provincia Autonoma di Trento and Q@TN, the joint laboratory between University of Trento, FBK—Fondazione Bruno Kessler, INFN—National Institute for Nuclear Physics, and CNR—National Research Council.
S.S.\ and L.E.\ acknowledge funding by the European Union under Horizon Europe Programme, Grant
Agreement 101080086–NeQST.

\section*{Author contribution}
S.S.\ and X.B-M.\ conceived the project.
S.S.\ performed the experiments on the IBM quantum hardware platform.
X.B-M., and L.E.\ performed the classical numerical simulations to compare to the experimental data.
A.B.\ performed the density matrix numerical simulations.
All authors analyzed the data, interpret the results, and wrote the manuscript.

\section*{Competing interests}
The authors declare no competing interests.

\section*{Data availability}
The data supporting the findings of this study are available from the corresponding author upon reasonable request.

\section*{Code availability}
All packages used in this work are open source and available via git repositories online.

\end{document}